\documentclass{aa}
\usepackage[varg]{txfonts}
\usepackage{natbib}
\bibpunct{(}{)}{;}{a}{}{,}
\usepackage[colorlinks=true, linkcolor=blue, citecolor=blue, filecolor=blue, urlcolor=blue]{hyperref}
\usepackage{pgf}
\usepackage{pgfplotstable}
\usepackage{amsmath}
\usepackage{comment}
\usepackage{ulem}
\usepackage{CJKutf8}

\newcommand{\tktext}[1]{\begin{CJK}{UTF8}{mj}#1\end{CJK}}

\begin{document}

\title{A comparison of particle and fluid approaches for modeling the streaming instability in global patch simulations}

\author{Dominik Hetzel Ostertag \inst{\ref{inst1}, \ref{inst2}} \corrauth{ostertag@mpia.de}\and 
Mario Flock \inst{\ref{inst1}} \email{flock@mpia.de}\and 
Paul Rosendahl \inst{\ref{inst1}} \email{info@paul-rosendahl.de}\and 
Andrea Mignone \inst{\ref{inst3}} \email{andrea.mignone@unito.it}\and 
Jeonghoon Lim (\tktext{임정훈}) \inst{\ref{inst4}, \ref{inst5}, \ref{inst6}} \email{jay.lim1@unlv.edu}\and 
Jacob B. Simon \inst{\ref{inst4}} \email{jbsimon@iastate.edu}}
\authorrunning{D. Hetzel Ostertag et al.}

\institute{Max-Planck-Institut für Astronomie, Königstuhl 17, 69117 Heidelberg, Germany \\ \email{ostertag@mpia.de} \label{inst1} 
\and
Fakultät für Physik und Astronomie, Universität Heidelberg, Im Neuenheimer Feld 226, 69120 Heidelberg, Germany \label{inst2}
\and
Dipartimento di Fisica, Università degli Studi di Torino, Via Pietro Giuria 1, I-10125 Torino, Italy \label{inst3}
\and 
Department of Physics and Astronomy, Iowa State University, Ames, IA 50010, USA \label{inst4}
\and
Nevada Center for Astrophysics, University of Nevada, Las Vegas, 4505 South Maryland Parkway, Las Vegas, NV 89154, USA
\label{inst5}
\and
Department of Physics and Astronomy, University of Nevada, Las Vegas, 4505 South Maryland Parkway, Las Vegas, NV 89154, USA
\label{inst6}}

\abstract
{Identifying the process relevant to planet formation is still a major task. 
Mechanisms have been identified that can trap dust and prevent it from migrating inward. 
A particular hydrodynamic instability, known as the streaming instability, is of high interest because it might cause the formation of dust clumps from which the first planetesimals form.}
{We compared the outcome of nonlinear streaming instability simulations using two fundamental approaches in a more realistic global disk model. 
The results were not used to identify or disentangle the effect of numerical diffusion in fluid simulations, for instance, or the possibility of crossing particle streams. 
Instead, they enabled us to determine whether the Lagrangian particle approach is most suitable for numerical studies of the streaming instability.}
{We performed 2D axisymmetric global and isothermal hydrodynamic simulations using the two codes FARGO3D (dust fluid) and PLUTO (fluid and particles) to investigate the temporal evolution of dust scale height, the midplane dust-to-gas ratio, and the maximum dust density, as well as to study velocity distributions and vorticities. 
For PLUTO particle simulations, we also investigated the influence of the gas Riemann solver and spatial reconstruction orders.}
{Our results from a resolution study of up to 5120 cells per gas scale height indicate convergence between PLUTO fluid and FARGO3D. 
The two codes yield a Gaussian profile that is dominated by small-scale dust-depleted voids. 
The particle simulations with PLUTO have not yet converged because a transient midplane dust component persists, indicating the need for an even higher resolution or more particles. 
A more accurate Riemann solver or a higher-order spatial reconstruction scheme helps to cause the dust density profile to approach the converged structure. 
At the highest resolution, the maximum dust density in fluid simulations is only 36$\%$ of that in particle simulations, which is similar to other code-comparison works.}
{When the streaming instability is studied in a global sense, the choice of the Riemann solver and spatial reconstruction order becomes more important than the approach itself (fluid or particles). 
When the computational performance of the two dust treatments is included, the fluid approach provides an efficient and suitable tool for streaming instability simulations.}

\keywords{Accretion, accretion disks - Instabilities - Methods: numerical - Planets and satellites: formation - Protoplanetary disks }

\titlerunning{Testing particle and fluid approaches for global streaming instability simulations}
\maketitle
\nolinenumbers

\section{Introduction}
It is crucial to understand the properties of dust grains embedded in protoplanetary disks to understand how planets form. 
Recent observations on protoplanetary disks with the Atacama Large Millimeter Array (ALMA), such as \cite{Sierra2025}, \cite{Han2023}, and \cite{DoiKataoka2023}, highlighted that the largest available dust grains are on the order of millimeters or centimeters. 
On the other hand, observations of the disk around PDS 70 showed clear indications of already-formed giant planets \citep[]{KepplerM2018, Trevascus2025}. 
This raises the question of how efficient and rapid growth by many orders of magnitude in grain size is possible to form the first larger objects, such as planetesimals. 
In recent years, the streaming instability \citep[][]{YoudinGoodman2005} has gained considerable focus, as reported by \cite{JohansenYoudin2007} and \cite{JohansenOishiNature2007}, who noted that the streaming instability can form overdensities that can then collapse gravitationally to form planetesimals that are typically on the order of one to hundreds of kilometers in diameter \citep[see, e.g.,][]{JohansenYoudinMacLow2009, Nesvorny2021}. 
\cite{Hughes2018} emphasized that observable debris disks are an indication that planet formation processes have successfully produced enough planetesimals that are used as the building blocks for planets. 
Many works, such as \cite{JohansenYoudinMacLow2009}, \cite{SimonArmitageLiYoudin2016}, \cite{KlahrSchreiber2020}, and \cite{SchaeferJohansen2024}, have studied planetesimal formation via streaming instability, showing that when dust is sufficiently concentrated, many orders of magnitude in grain size can be skipped. 
\cite{PolakKlahr2023} have studied the formation of individual planetesimals through the gravitational collapse of a single pebble cloud at high resolution. This provided further insights into planetesimal formation. 

Based on certain conditions for the Stokes number and the global dust-to-gas content in the disk, works such as \cite{CarreraD2015}, \cite{Yang2017}, \cite{Li2021}, or recently \cite{LimSimon20253D} have investigated whether these clumps can grow in density until they reach a critical density threshold and collapse due to their own gravity. 
These works showed that certain combinations allow dense clumps to form, and that these combinations therefore provide thresholds for simulations that cannot resolve streaming instability. 
Typically, shearing box simulations in combination with the particle approach for dust are used, which do not allow us to study the instability on global scales. 
The work of \cite{Schäfer2020}, subsequent studies, and the work of \cite{HuangBai2025} have begun to investigate the streaming instability in conjunction with vertical shear instability (VSI) in global disk models, spanning several tens of astronomical units in size. 
They demonstrated that the two instabilities interact, which might affect the criteria for the formation of dense clumps. 

When the streaming instability is studied in a global setup with high resolution (as in \cite{Flock2021} with PLUTO and Lagrangian dust particles and in \cite{Ostertag2025StrongClumping} with FARGO3D), the fluid approach and the particle approach for dust allow us to study this instability and its clumping behavior. 
This raises the question of how the two approaches compare and what their respective strengths and limitations are in capturing the streaming instability and its clumping behavior.

A first comprehensive work that compared the streaming instability across different hydrodynamical codes and dust approaches was provided by \cite{Baronett+2026Pfits} using unstratified shearing box simulations.
We compare the particle and fluid approaches for dust in global models with vertical stratification, similar to those in \cite{Flock2021}. 
However, we also compare the effect of different gas Riemann solvers and spatial reconstruction schemes, as well as the difference between two different hydrodynamics codes (PLUTO and FARGO3D).

The structure of the paper is as follows: First, we describe in Sect. \ref{sec: Methods} the disk model, followed by an investigation of the effect of different reconstruction schemes on simulations with dust particles and a fixed resolution in Sect. \ref{sec: reconstruction_schemes}. 
Next, we compare in Sect. \ref{sec: riemann_solver} three different and well-known Riemann solvers, again for dust particle simulations with a fixed resolution. 
In Sect. \ref{sec: pluto_fluid_pluto_particles} we compare PLUTO simulations with dust fluid and dust particles, and in Sect. \ref{sec: pluto_fluid_fargo3d}, we compare simulations between PLUTO and FARGO3D, both with a dust fluid. 
After investigating the convergence of particle and fluid setups with respect to the dust scale height, velocities, and vorticity, we compute the performance of the different hydro codes and numerical setups. 
We conclude this work with a discussion and a conclusion in Sects. \ref{sec: Discussion} and \ref{sec: Conclusion}, respectively.

\section{Methods}
\label{sec: Methods}

\subsection{Disk setup}
\label{sec: disk setup}
In the following, we briefly describe our disk setup, which used the steady~state solution of an unperturbed disk from \cite{Nakagawa1986} as the initial condition. 
For simplicity, we set the initial vertical velocities of dust and gas, $V_{\rm z}$ and $v_{\rm z}$, respectively, to zero throughout the disk. 
The radial and azimuthal velocity components of dust ($V_{\rm R}$, $V_{\phi}$) and gas ($v_{\rm R}$, $v_{\phi}$) were initialized with the following expressions in cylindrical geometry ($R$, $\phi$, $z$):
\begin{gather} 
    \label{eq: radial dust velocity}
    V_{\rm R} = -\frac{\rho_{\rm g}}{\rho_{\rm g} + \rho_{\rm d}}\frac{2 D \Omega_{\rm K}}{D^2 + \Omega^2_{\rm K}} \eta R \Omega_{K}, \\
    V_{\phi} = \left(1 -\frac{\rho_{\rm g}}{\rho_{\rm g} + \rho_{\rm d}}\frac{D^2}{D^2 + \Omega^2_{\rm K}} \eta \right) R \Omega_{K}, \\
    v_{\rm R} = \frac{\rho_{\rm d}}{\rho_{\rm g} + \rho_{\rm d}}\frac{2 D \Omega_{\rm K}}{D^2 + \Omega^2_{\rm K}} \eta R \Omega_{K}, \\
    v_{\phi} = \left[1 + \left(\frac{\rho_{\rm d}}{\rho_{\rm g} + \rho_{\rm d}}\frac{D^2}{D^2 + \Omega^2_{\rm K}} -1 \right) \eta \right] R \Omega_{K}.
\end{gather}
The variables $\rho_{\rm g}$ and $\rho_{\rm d}$ describe the gas and dust volume densities, respectively. 
$\Omega_{\rm K} = \sqrt{G M_{*} / R^3}$ is the Keplerian angular velocity, where $G$ is the gravitational constant, and $M_{*}$ is the mass of the central star. 
$D$ was computed with
\begin{equation}
    D = \frac{1+\epsilon}{\mathit{St}} \Omega_{K},
\end{equation}
where $\epsilon = \rho_{\rm d} / \rho_{\rm g}$ describes the dust-to-gas ratio and $\mathit{St}$ is the dimensionless Stokes number (in our models, a global constant). $\eta$ is linked to the radial pressure gradient with
\begin{equation}
    \eta = - \frac{1}{2 \rho_{\rm g} R \Omega^2_{\rm K}} \frac{\partial P}{\partial R}.
\end{equation}
As we treated the gas as a locally isothermal fluid, we defined the pressure as $P=\rho_{\rm g} c^2_{\rm s}$ and the sound speed, $c_{\rm s}(R)$, was only a function of radius. 
As a consequence, the vertical pressure gradient was entirely determined by the vertical density stratification. 
In our disk model, the gas and dust volume densities were given by
\begin{equation}
    \rho_{g}(R, z) = \rho_{g, 0} \left(\frac{R}{R_{\rm 0}} \right)^P \exp \left[\frac{R^2}{H^2_{\rm g}} \left( \frac{R}{r} - 1 \right) \right],
    \label{eqn: gasdensity}
\end{equation}
where $r=\sqrt{R^2 + z^2}$ is the spherical radius, and we computed the gas scale height with
\begin{equation}
    H_{\rm g} = H_{\rm g,0} \left( \frac{R}{R_{\rm 0}} \right)^{\frac{Q+3}{2}}.
    \label{eqn: gas scale height}
\end{equation}
For dust, we used similar expressions for density and scale height as given in Eq. \ref{eqn: gasdensity} and \ref{eqn: gas scale height}, but replaced the midplane gas density $\rho_{g, 0}$ with the midplane dust density $\rho_{d, 0}$ and the gas scale height at $R_{\rm 0}$ with the dust scale height $H_{\rm d, 0}$. 
The gas and dust midplane densities, $\rho_{\rm g, 0}$ and $\rho_{\rm d, 0}$, were then computed with
\begin{equation}
    \rho_{0} = \frac{\Sigma_{0}}{\sqrt{2\pi} H_{\rm 0}}.
    \label{eq: midplane density}
\end{equation}
An overview of parameter values is given in Table \ref{tab: disk setup}.
\begin{table}[ht]
\caption{Disk setup values.}
\label{tab: disk setup}
\centering
\begin{tabular}{l l | l l}
\hline\hline
Parameter & Value & Parameter & Value\\
\hline
$\theta_{\rm min}$      &  1.556796     & $r_{\rm min}$             & 9.3                \\
$\theta_{\rm max}$      &  1.584796     & $r_{\rm max}$             & 10.7               \\
$R_{0}$                 &  10 au        & $\Sigma_{\mathrm{g},0}$   & 60 $\rm gcm^{-2}$  \\
$H_{\rm g, 0} / R_{0}$  &  0.07         & $\Sigma_{\mathrm{d},0}$   & 0.6 $\rm gcm^{-2}$ \\
$H_{\rm d, 0} / R_{0}$  &  0.0014       & $\mathrm{St}$               & 0.01               \\
$\Pi$                   &  0.07         & $M_{*}$                   &  $M_{\odot}$       \\
$\eta$                  &  0.0049       & $P$, $Q$                  & -1, -1             \\
\hline
\end{tabular}
\tablefoot{The parameter $\Pi= \eta v_{\rm K} / c_{\rm s}$ is a dimensionless quantity that determines the deviation from pure Keplerian velocity due to the background pressure gradient relative to the sound speed (see Sect. 2 of \cite{SekiyaOnishi2018}). The Stokes number is a global constant in our models, making the grain size a function of location in the disk.}
\end{table}

\subsection{Numerical setups}
\label{sec: numericalSetups}
We performed 2D, axisymmetric, and isothermal simulations using spherical coordinates in meridional and radial directions. 
The simulations were done with the PLUTO\footnote{\url{https://plutocode.ph.unito.it/} (Version 4.4-patch3)} \citep{Mignone2007} and the FARGO3D\footnote{\url{https://fargo3d.github.io/documentation/} (public commit 22.7.2020)} \citep{Benítez-Llambay2016} hydrodynamics codes.
PLUTO enabled us to simulate dust as a pressureless fluid and as Lagrangian particles, whereas we used FARGO3D to model dust as a pressureless fluid. 
In all PLUTO simulations, we did not consider explicit gas viscosity, background turbulence, or dust diffusion. 
For the FARGO3D simulations, the $\alpha$-viscosity \citep[for $\alpha$ see][]{Shakura_Sunyaev_1973} and diffusion module \citep[see][]{Weber2019} were used. 
The latter can generally mimic turbulent dust diffusion driven by instabilities such as the magneto-rotational instability or the VSI, using the same $\alpha$-parameter. 
We set the value of $\alpha$ to zero for all FARGO3D simulations to study the strength of turbulence solely through streaming instability or other dust-gas instabilities, such as the Kelvin-Helmholtz instability (KHI). 

To simulate dust particles using PLUTO, we used the dust particle module introduced in \cite{Mignone2019DustParticleModule}. 
In all particle simulations, we used the exponential midpoint method to evolve dust particles in time, while gas quantities were evolved in time using a second-order Runge-Kutta scheme (RK2). 
We used the Cloud-In-Cell particle shape function to interpolate gas quantities onto the particle positions. 
Unless otherwise stated, all particle simulations used $1.04\cdot10^{7}$ Lagrangian particles, independent of the underlying resolution. 
This number is the same as used in \cite{Flock2021}. 
For our model with a resolution of $2048\times512$, this corresponds to 9.9 particles per cell. 
For the setup with a resolution of $4096\times1024$, we additionally performed a simulation with four times as many particles ($4.17\cdot10^{7}$) to investigate the sensitivity of dust properties on the particle resolution. 
We set the Courant–Friedrichs–Lewy (CFL) value to 0.2 for all particle simulations. 
An overview of the numerical combinations used for spatial reconstruction, the gas Riemann solver, and slope limiters is provided in Table \ref{tab: simulation overview} (see \texttt{conf01} - \texttt{conf05}).
\begin{table*}
\caption{Overview of configurations.}
\label{tab: simulation overview}
\centering
\begin{tabular}{l c c c c c c c}
\hline\hline
Name        & Resolution                        &  Vert. domain  & Reconstruction   &  Time stepping    & Gas Riemann   & Flux limiter   & CFL value  \\
            & $N_{\rm r} \times N_{\theta}$     &   size         &                  &                   & solver    &                &            \\
\hline \\ [-1.0ex]
\multicolumn{8}{c}{PLUTO Particles} \\ 
\hline 
\texttt{conf01}      &  $4096\times512$       &   0.2 $H_{\rm g}$  &  PLM              &  RK2              & HLL     &   MC Limiter      & 0.2           \\
\texttt{conf02}      &  $4096\times512$       &   0.2 $H_{\rm g}$  &  PPM-4            &  RK2              & HLL     &   -               & 0.2           \\
\texttt{conf03}      &  $4096\times512$       &   0.2 $H_{\rm g}$  &  PLM              &  RK2              & HLLC    &   MC Limiter      & 0.2           \\
\texttt{conf04}      &  $4096\times512$       &   0.2 $H_{\rm g}$  &  PLM              &  RK2              & ROE     &   MC Limiter      & 0.2           \\
\texttt{conf05}      &  varying               &   0.4 $H_{\rm g}$  &  PPM-4            &  RK2              & ROE     &   -               & 0.2           \\
\hline \\ [-1.0ex]
\multicolumn{8}{c}{PLUTO Fluid} \\ 
\hline
\texttt{conf06}      & varying                &   0.4 $H_{\rm g}$  &  PLM              &  RK2              & HLL     &  van Leer Limiter & 0.2           \\
\texttt{conf07}      & varying                &   0.4 $H_{\rm g}$  &  PPM-4            &  RK3              & ROE     &   -               & 0.2           \\
\hline \\ [-1.0ex]
\multicolumn{8}{c}{FARGO3D Fluid} \\ 
\hline
\texttt{conf08}      & varying                &   0.4 $H_{\rm g}$  &  PLM              &  RK2              & -       &  van Leer Limiter & 0.44          \\
\hline\hline   
\end{tabular}
\tablefoot{PLM stands for piece-wise linear, which is second-order accurate. 
PPM-4 stands for piece-wise parabolic, which is fourth-order accurate in this work. 
HLL is an abbreviation for the Harten, Lax, and van Leer solver, while MC limiter is the monotonized central limiter. In PLUTO, the slope limiter is only used when the spatial reconstruction is PLM. 
The information on spatial reconstruction and slope limiter for FARGO3D was taken from Sect. 3.4 in \cite{Benítez-Llambay2016}. 
The information "varying" in column 2 indicates that for this configuration, we studied the disk for increasing resolutions, starting at a resolution of $512\times128$ and increased the resolution in each direction by factors of 2 until a final resolution of $4096\times1024$ for \texttt{conf06} and $8192\times2048$ for the setups \texttt{conf05}, \texttt{conf07}, and \texttt{conf08}.}
\end{table*}

To model dust as a fluid, we used the same physical setup (Eq. \ref{eq: radial dust velocity} - \ref{eq: midplane density}). 
For PLUTO simulations, we used the dust fluid module recently published by \cite{Ziampras2025DustFluidModule}. 
We used an exact solver \citep{Leveque2004} to compute cell interface fluxes for the dust fluid, while the dust was evolved in time using a second-order accurate explicit time integration method. 
Table \ref{tab: simulation overview} lists the numerical configuration for two setups (\texttt{conf06} for a numerically more diffusive setup and \texttt{conf07} for a numerically more accurate setup). 
As with the particle simulations, we also employed a CFL value of 0.2.
To be able to simulate dust as a fluid using FARGO3D, we employed the numerical methods for systems consisting of gas and potentially many dust fluid species as described in \cite{Benítez-Llambay2019}, but we considered only one dust species in this work. 
The setup was already described in \cite{Ostertag2025StrongClumping}, where we used a CFL value of 0.44.

At this point, we stress that the considered vertical domain size in \cite{Flock2021} appears to be too small for PLUTO and FARGO3D simulations. 
We show in Sect. \ref{sec: influenceDomainSize} that a small vertical domain size can lead to an artificial increase in the dust scale height. 
For this reason, the vertical domain size in \cite{Ostertag2025StrongClumping} was already increased by a factor of 2 to 0.4 $H_{\rm g}$ for FARGO3D simulations, while keeping the resolution fixed. 
In Sects. \ref{sec: reconstruction_schemes} and \ref{sec: riemann_solver}, where we analyze the influence of different spatial reconstruction schemes and different Riemann solvers, we used the original vertical domain size as in \cite{Flock2021} of 0.2 $H_{\rm g}$. 
However, in the following sections, where we directly compare fluid and particle simulations, we increased the vertical domain of PLUTO simulations while maintaining a fixed resolution. 
For better differentiation, the information on the size of the entire vertical domain is included in Table \ref{tab: simulation overview}. 
In addition, we mark simulations in Table \ref{tab: overview} whose vertical domain is not increased to 0.4$\,H_{\rm g}$ with a dagger symbol.

\subsection{Boundary conditions}
\label{sec: boundary conditions}
In the following, we briefly describe the boundary conditions we applied. 
Particles were simply removed from the simulation once they crossed a boundary (the inner or the outer) in any of the directions and left the computational domain. 
For gas and when dust is treated as a fluid, the following conditions in PLUTO (FARGO3D) were used: in the radial direction, the gas and dust hydro variables had zero gradients (symmetric boundary conditions), but neither gas nor dust were allowed to flow into the computational domain. 
This was achieved by setting the radial velocity at the radial interface between the computational domain and the boundary to zero if the velocity pointed into the domain. 
In the meridional direction, we extrapolated the exponential gas density profile with a zero gradient (using symmetric boundary conditions for the dust density). 
For the gas (dust) velocities, we again used a zero-gradient (symmetric boundary conditions) approach to prevent gas and dust from flowing into the computational domain. 
See also Sect. 2.1 in \cite{Flock2021} and Appendix A in \cite{Ostertag2025StrongClumping}.

\subsection{Damping zones}
\label{sec: damping zones}
In all dust-fluid and particle simulations, we damped the gas density in the entire domain back to the initial condition to prevent gas mass loss, following \cite{Flock2021} (see their Sect. 2.3 and Eq. 15). 
We used the same damping timescale of 1$\Omega^{-1}$, and $\Omega$ evaluated at 1 astronomical unit (AU). 
As \cite{Flock2021} have shown in their Appendix C, the influence of the damping timescale on the gas density is small in this regime. 
Additionally, we added damping or buffer zones at the inner and outer domain edges: For all PLUTO simulations, we applied Eq. 14 from \cite{Flock2021} (with $R_{\rm b}=9.3$ AU, $R_{\rm e}=10.7$ AU, and $ w=0.05$), while for all FARGO3D simulations we refer to Appendix B and Eq. B.4 in \cite{Ostertag2025StrongClumping} ($r_{\rm inf}=9.349$, $r_{\rm sup}=10.643$). 
Across all simulations, we damped the gas's azimuthal and radial velocities in the buffer zones to remove oscillations near the boundary. 
The gas relaxation in density and velocities happened on the same timescale of 1$\Omega^{-1}$, to be consistent with what was used in \cite{Flock2021}. 
In each simulation, we reduced the azimuthal velocity of dust in the inner damping zone at each timestep with a discrete numerical integration step of the form $v^{n+1}_{\mathrm{d, \phi}} = (1-10^{-8}) \cdot v^{n}_{\mathrm{d, \phi}}$ to artificially increase the radial velocity and prevent material accumulation. 
The vertical velocity of dust is damped in the outer buffer zone using a similar numerical integration step of the form $v^{n+1}_{\mathrm{d, \theta}} = (1-10^{-5}) \cdot v^{n}_{\mathrm{d, \theta}}$ for PLUTO simulations, while we used Eq. B3 from \cite{Ostertag2025StrongClumping} for FARGO3D simulations, also with a timescale of $T_{\rm orb}(1\,\rm au)/(2\pi)$.

Dust particles were refilled at the outer damping zone, as stated in Sect. 2.4 in \cite{Flock2021}, and the dust fluid in PLUTO and FARGO3D was refilled as stated in Appendix B in \cite{Ostertag2025StrongClumping}: The dust density was set back to the initial condition at every timestep, once a cell inside the damping zone reached a dust density of 25\% of the initial value.

Whenever we computed the dust-to-gas ratio, we used the local gas density at the same output as the dust density, and we considered only the radial domain 10 $\pm$ 0.15 AU to compute the average. 
To study the evolution of the maximum dust density, we only considered the radial domain from 9.4 to 10.4 AU to exclude the effects of the outer radial damping zone, in which dust was refilled, as well as the outer radial boundary conditions.

\subsection{Dust scale height}
\begin{figure}
    \centering
    \includegraphics[]{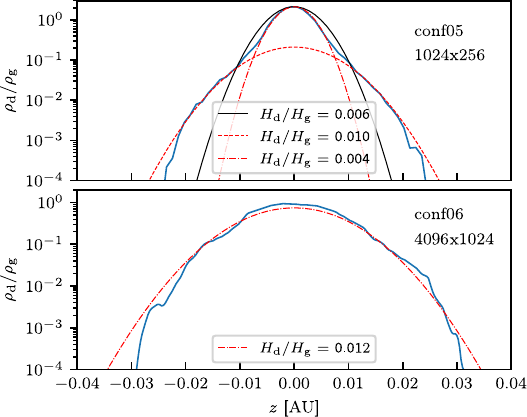}
    \caption{Determination of the vertical dust scale height for two different numerical setups at 100 orbits for a resolution of $4096\times1024$. 
    The vertical dust-to-gas ratio profile (averaged between 10 $\pm$ 0.15 AU) of the models is shown in blue. 
    \textit{Top}: An example of the particle setup \texttt{conf05}, where the dashed and dash-dotted lines show the midplane and broad component, respectively. 
    The solid black line shows the Gaussian profile that we obtained with the dust scale height estimated by Eq. \ref{eqn: ComputeDustScaleHeight}. 
    \textit{Bottom}: An example for the fluid setup \texttt{conf06}, where the dash-dotted line shows the best-fit profile that we obtained with a simple Gaussian.}
    \label{fig: demonstrationDustScaleHeightComputationFluid}
\end{figure}

To facilitate a quantitative comparison between different numerical approaches, we determined the dust scale height in the vertical direction. 
Throughout this work, we encounter two distinct dust density profiles that depend on the specific numerical setup. 
The first one shows a rather typical Gaussian shape, while the second one can be described as a combination of two individual Gaussians, one for a midplane component and one for the broad dust distribution. 
See Fig. \ref{fig: demonstrationDustScaleHeightComputationFluid} for two example setups. 

\cite{Yang2017} provided a method to estimate the dust scale height, in which the ratio of the scale heights of dust and gas is given by the standard deviation of the vertical position of dust grains, $z_p$:
\begin{equation}
    \frac{H_{\mathrm{p}}}{H} = \frac{\sqrt{\overline{z_p^2} - \overline{z_p}^2}}{H},
    \label{eqn: ComputeDustScaleHeight}
\end{equation}
with $z_p= r_p \cdot \cos(\theta_p)$. 
The horizontal bar shows the quantities over which averaging is performed. 
For the PLUTO dust particle simulations, this approach is straightforward, as each particle has its own unique position. 
For the setup \texttt{conf05}, the vertical dust-to-gas ratio (shown in blue) and the expected Gaussian profile using Eq. \ref{eqn: ComputeDustScaleHeight} (black) are shown in the top panel of Fig. \ref{fig: demonstrationDustScaleHeightComputationFluid}. 
For this setup, Eq. \ref{eqn: ComputeDustScaleHeight} underestimates the dust scale height of the broad component and yields a scale height between the midplane and the broad component. 
This means that this method is only suitable if there is no additional midplane component. 
Therefore, we decided to fit a Gaussian function to each component, and the resulting best-fit profiles are also shown in the top panel of Fig. \ref{fig: demonstrationDustScaleHeightComputationFluid} as a dashed and dash-dotted red line. 
For the fluid simulations (performed with either PLUTO or FARGO3D), Eq. \ref{eqn: ComputeDustScaleHeight} cannot be applied, simply because we do not have individual particle locations. 
Instead, we also fitted a Gaussian profile to each density component. 
To obtain meaningful results, we applied specific dust-to-gas ratio thresholds when fitting the Gaussians to the simulation data. 
To determine the broad dust scale height, we used dust-to-gas ratio values of $\leq 0.03$ for \texttt{conf01}–\texttt{conf04} and $\leq 0.04$ for \texttt{conf05}. 
For the PLUTO fluid setup \texttt{conf06}, the threshold was increased to 0.1.
For the PLUTO fluid and FARGO3D simulations (\texttt{conf07} and \texttt{conf08}, respectively), we considered only values $\leq$0.02. 
Because we did not know the amplitude of the broad density component and could only fit its wings, we allowed the fit amplitude to vary between 0 and 6. 
This can result in variations in the values of the dust scale height.  
The midplane component for the setups \texttt{conf01} - \texttt{conf05}, \texttt{conf07}, and \texttt{conf08} was determined using dust-to-gas ratios $\geq$ 0.4. 
It should be noted that at a certain height above the midplane, where the density starts to drop significantly, the fit no longer accurately describes the distribution.

\section{Results}
In Table \ref{tab: overview} we provide an overview of values for characteristic parameters across all dust fluid and dust particle simulations. 
For all parameters, we calculated a temporal average starting at 30 orbits and continued until the end of the simulations. 
For simulations with a resolution lower than or equal to $4096\times1024$, the end of the simulation is reached after 100 orbits. 
For simulations with a resolution of $8192\times2048$, the end was chosen to be at 80 orbits. 
The simulations will be presented in more detail in the following sections.
\begin{table}
\caption{Overview of midplane dust-to-gas ratios, maximum dust densities, and dust scale height for the broad and midplane component during the saturated state of the streaming instability for all simulations.}
\label{tab: overview}
\centering
\begin{tabular}{l c c c c}
\hline\hline
Setup   &  $\left. \rho_{\rm d} / \rho_{\rm g} \right|_{z=0}$  &  \rm max($\rho_{\rm d}$)  & $\left. H_{\rm d} / H_{\rm g} \right|_{B}$  & $\left. H_{\rm d} / H_{\rm g} \right|_{M}$ \\
        &                                                      &  [$\mathrm{gcm^{-3}}$]    &                                             &                                            \\
\hline \\ [-1.0ex]

\multicolumn{5}{c}{PLUTO Particles} \\
\hline
    \texttt{conf01}$\dagger$    & 0.78 & 3.83 $\cdot 10^{-12}$ & 0.0150 & - \\
    \texttt{conf02}$\dagger$    & 1.06 & 6.45 $\cdot 10^{-12}$ & 0.0138 & 0.0079 \\
    \texttt{conf03}$\dagger$    & 1.07 & 7.85 $\cdot 10^{-12}$ & 0.0155 & 0.0075 \\
    \texttt{conf04}$\dagger$    & 1.08 & 7.65 $\cdot 10^{-12}$ & 0.0156 & 0.0072 \\
\\ [-1.0ex]

\multicolumn{5}{c}{PLUTO Particles - \texttt{conf05}} \\
\hline
    $512\times128$             & 2.56 & 4.50 $\cdot 10^{-11}$ & 0.0076 & 0.0033 \\
    $1024\times256$            & 2.04 & 1.61 $\cdot 10^{-11}$ & 0.0096 & 0.0046 \\
    $2048\times512$            & 1.61 & 1.06 $\cdot 10^{-11}$ & 0.0113 & 0.0055 \\
    $4096\times1024$           & 1.34 & 8.91 $\cdot 10^{-12}$ & 0.0118 & 0.0065 \\
    $8192\times2048$           & 0.96 & 6.68 $\cdot 10^{-12}$ & 0.0134 & 0.0093 \\
\\ [-1.0ex]

\multicolumn{5}{c}{PLUTO Fluids - \texttt{conf06}} \\
\hline
    $512\times128$             & 0.25 & 1.10 $\cdot 10^{-12}$ & 0.0044 & - \\
    $1024\times256$            & 0.47 & 4.13 $\cdot 10^{-12}$ & 0.0127 & - \\
    $2048\times512$            & 0.77 & 4.02 $\cdot 10^{-12}$ & 0.0123 & - \\
    $4096\times1024$           & 0.93 & 3.50 $\cdot 10^{-12}$ & 0.0122 & - \\
\\ [-1.0ex]

\multicolumn{5}{c}{PLUTO Fluids - \texttt{conf07}} \\
\hline
    $512\times128$             & 2.61 & 2.75 $\cdot 10^{-11}$  & 0.0048  & 0.0033 \\
    $1024\times256$            & 1.61 & 1.29 $\cdot 10^{-11}$  & 0.0118  & 0.0056 \\
    $2048\times512$            & 1.29 & 7.42 $\cdot 10^{-12}$  & 0.0120  & 0.0063 \\
    $4096\times1024$           & 1.10 & 6.35 $\cdot 10^{-12}$  & 0.0127  & 0.0076 \\
    $8192\times2048$           & 0.63 & 3.15 $\cdot 10^{-12}$  & 0.0129 &  -\\
\\ [-1.0ex]

\multicolumn{5}{c}{FARGO3D - \texttt{conf08}} \\
\hline
    $512\times128$             & 1.14 & 7.02 $\cdot 10^{-12}$  & 0.0127 & 0.0074 \\
    $1024\times256$            & 1.05 & 5.85 $\cdot 10^{-12}$  & 0.0122 & 0.0077 \\
    $2048\times512$            & 0.94 & 4.95 $\cdot 10^{-12}$  & 0.0126 & 0.0085 \\
    $4096\times1024$           & 0.68 & 2.48 $\cdot 10^{-12}$  & 0.0134 & - \\
    $8192\times2048$           & 0.51 & 2.39 $\cdot 10^{-12}$  & 0.0141 & - \\
\hline\hline
\end{tabular}
\tablefoot{Every value we list above is a temporal average from 30 orbits (the nonlinear streaming instability reached a saturated state) to the end of the simulation. 
Column 2 shows the midplane dust-to-gas ratio averaged in the radial domain 10 $\pm$ 0.15 AU. 
Column 3 shows the maximum dust density in units of $\rm gcm^{-3}$ and only considers values between 9.4 and 10.4 AU. 
Columns 4 and 5 are the dust scale height in units of the gas scale height for the broad and midplane component, respectively. 
The entries in the last column that show a minus sign indicate that the dust density shows no clear midplane component, and a corresponding fit was not meaningful. 
Simulations marked with a dagger used the original vertical domain size from \cite{Flock2021}, which influences the dust density. See Sect. \ref{sec: influenceDomainSize} for more information. 
The numerical configurations \texttt{conf01} - \texttt{conf08} are explained in more detail in Table \ref{tab: simulation overview}.}
\end{table}

\subsection{Comparison of spatial reconstruction schemes}
\label{sec: reconstruction_schemes}
\begin{figure*}
    \centering
    \includegraphics[]{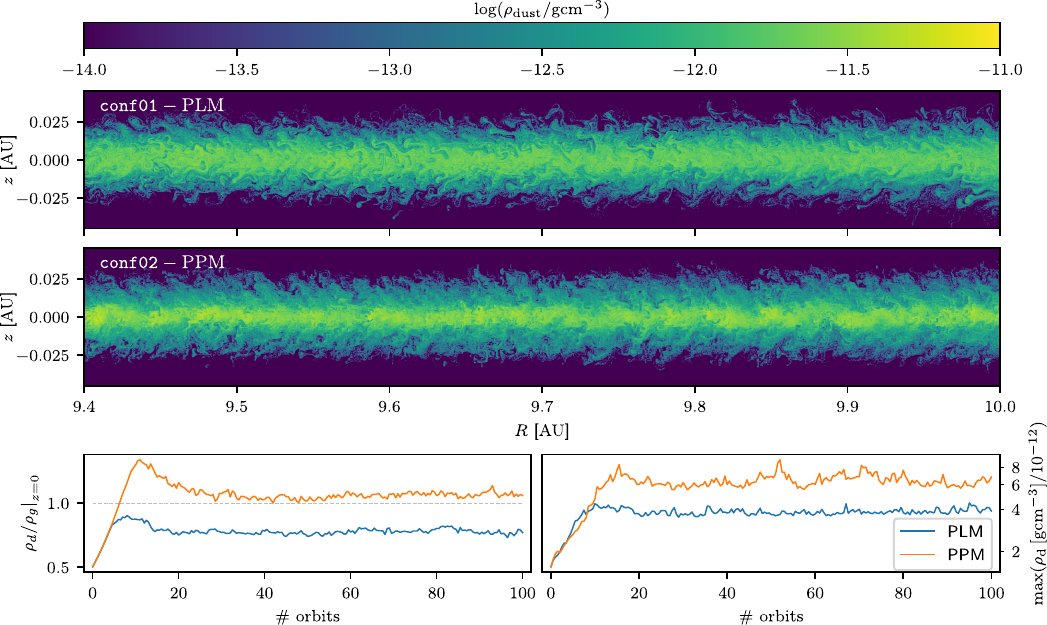}
    \caption{\textit{Top}: Comparison of the 2D dust density structure for the two numerical setups \texttt{conf01} and \texttt{conf02} after 100 orbits. 
    Both setups treated the dust as individual particles. 
    \textit{Bottom left}: The averaged dust-to-gas ratio in the region 10 $\pm$ 0.15 AU in the midplane. 
    The dashed~gray line indicates the location of equal gas and dust density. 
    \textit{Bottom right}: The maximum dust density in the radial domain [9.4, 10.4].}
    \label{fig: spatialReconstruction}
\end{figure*}

In this subsection, we aim to investigate the impact of a fourth-order piecewise parabolic scheme PPM (\texttt{conf02}) compared to a lower-order second-order piecewise linear scheme PLM (\texttt{conf01}) on the fundamental streaming instability properties, such as the midplane dust-to-gas ratio and the maximum dust density.
For this study, we used only PLUTO simulations in which dust was treated as discrete particles, with 4096 cells in the radial and 512 cells in the vertical direction. 
This corresponds to a resolution of 2560 cells per gas scale height. 
In Fig. \ref{fig: spatialReconstruction}, we show the 2D dust density of both simulations after 100 orbits (determined at 10 AU) as well as the temporal evolution of the midplane dust-to-gas ratio and the maximum dust density. 

First of all, we can see from the plots of the midplane dust-to-gas ratio or the maximum dust density that the nonlinear streaming instability has reached a steady state at {$\sim$} 30 orbits. 
In addition, the 2D dust density plots indicate that the dust density is structured by this nonlinear state. 
The comparison reveals that, for a linear reconstruction scheme (\texttt{conf01}), the vertical dust density profile can be approximated by a smooth single-Gaussian. 
In contrast, for a fourth-order parabolic scheme (\texttt{conf02}), an additional component is clearly visible in the midplane. 
Similar vertical density profiles for both scenarios are shown in Fig. \ref{fig: demonstrationDustScaleHeightComputationFluid}. 
At 100 orbits, the parabolic reconstruction scheme yields a dust-to-gas ratio exceeding 1.0 (compared to 0.77 for the linear reconstruction scheme) and a maximum density roughly twice that of the linear reconstruction. 

The dust scale height for both simulations, which we determined with Eq. \ref{eqn: ComputeDustScaleHeight}, is 0.014 $H_{\rm g}$ for a linear reconstruction scheme and 0.012 $H_{\rm g}$ for a parabolic scheme. 
Our obtained values for the dust scale heights can be compared with the results in \cite{LimSimon20252D}, who have already studied streaming instability in a 2D, axisymmetric, and stratified shearing-box model. 
Using the same set of parameters (constant Stokes number of 0.01, total dust-to-gas mass ratio of 0.01, vertical domain size of 0.2 $H_{\rm g}$, and vertical resolution of 2560/$H_{\rm g}$), the middle panel of their Fig. 5 shows a value around 0.009 $H_{\rm g}$, which is smaller than our obtained values. 

This is expected, as their value of $\sigma_{\rm d} = (2 \pi)^{1/2} Z/\Pi$ \citep{SekiyaOnishi2018} is 0.5 and slightly larger than our value of 0.36 (due to our larger value of $\Pi=0.07$), suggesting a higher dust scale height in our simulations.
By fitting a Gaussian profile to only the broad density component, we obtain slightly larger dust scale heights of 0.015 $H_{\rm g}$ and 0.0138 $H_{\rm g}$ for the linear and parabolic reconstruction, respectively.
As the vertical profile for a parabolic scheme shows a second, midplane component of approximately Gaussian shape, we determine a dust scale height of 0.008 $H_{\rm g}$. 
\cite{LimSimon20252D} also find the additional midplane component for a resolution of 1280/$H_{\rm g}$ to 5120/$H_{\rm g}$ (see, for instance, the left-hand side of their Fig. 6 as well as Appendix \ref{sec: athena}), but this component has almost disappeared for their highest resolution.

\subsection{Effect of the Riemann solver}
\label{sec: riemann_solver}
\begin{figure*}
    \centering
    \includegraphics[]{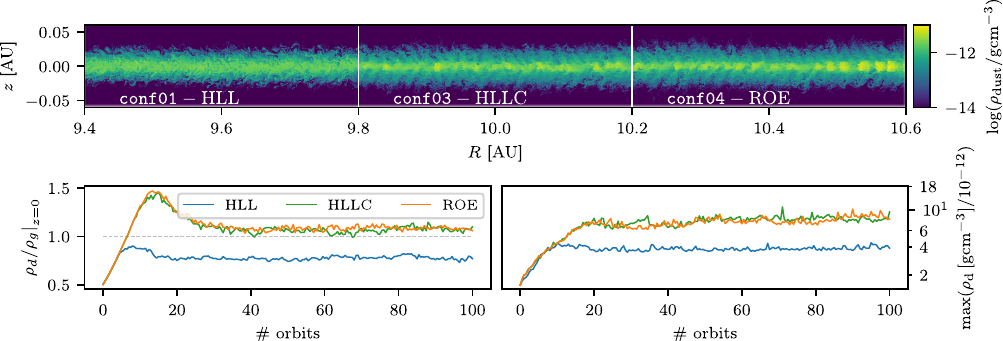}
    \caption{\textit{Top}: Comparison of the 2D dust density structures for three different Riemann solvers. 
    Each simulation is plotted after 100 orbits (at 10 AU). 
    The vertical white lines indicate the boundary between simulations with different Riemann solvers. 
    \textit{Bottom left}: The evolution of the average midplane dust-to-gas ratio in the region 10 $\pm$ 0.15 AU for each simulation with the number of orbits. 
    \textit{Bottom right}: The evolution of the maximum dust density in the radial regime [9.4, 10.4] AU.}
    \label{fig: 3RiemannSolvers}
\end{figure*}

For the second test, we aim to assess the extent to which the accuracy of the chosen gas Riemann solver affects the simulation outcomes. 
We again employed the particle approach with PLUTO, fixing the resolution to 4096 cells in the radial direction and 512 cells in the meridional direction. 
Again, this corresponds to a resolution of $2560/H_{\rm g}$. 
The three different Riemann solvers and numerical setups we test here are (listed with increasing accuracy of the solver): HLL (\texttt{conf01}), HLLC (\texttt{conf03}), and ROE (\texttt{conf04}). 
For HLL, the simulation is identical to the one in Fig. \ref{fig: spatialReconstruction}.
In Fig. \ref{fig: 3RiemannSolvers} we show the 2D dust density structure after 100 orbits for the three solvers, as well as the temporal evolution of the average midplane dust-to-gas ratio and the maximum dust density. 

For the HLLC and ROE solvers, we again observe an enhancement of dusty material in the midplane. 
As we only change the Riemann solver, this additional midplane dust enrichment must be related to fundamental properties of the solvers themselves in an isothermal case, where there exists no contact wave. 
For HLLC and ROE, we determine a dust scale height of the broad density component of 0.0155 $H_{\rm g}$ and 0.0156 $H_{\rm g}$, respectively. 
These values are about 12$\%$ and 13$\%$ larger than those obtained by using only a higher-order spatial reconstruction. 
For the midplane component, we obtain values of 0.0075 $H_{\rm g}$ and 0.0072 $H_{\rm g}$ for HLLC and ROE, respectively, which are 5$\%$ and 9$\%$ smaller than for a higher-order spatial reconstruction. 

As HLLC and ROE show an additional midplane component, the average dust-to-gas ratio increases to values close to 1.1, indicating a similar effect to that observed with higher-order spatial reconstruction. 
For the maximum dust density inside the radial domain between 9.4 and 10.4 AU, HLLC and ROE show similar values that can even reach values of $10^{-11}$ $\rm gcm^{-3}$. 
These maximum values are slightly larger than those for the fourth-order spatial reconstruction simulation (\texttt{conf02}) and exhibit an overall upward trend.

\begin{figure}
    \centering
    \includegraphics[]{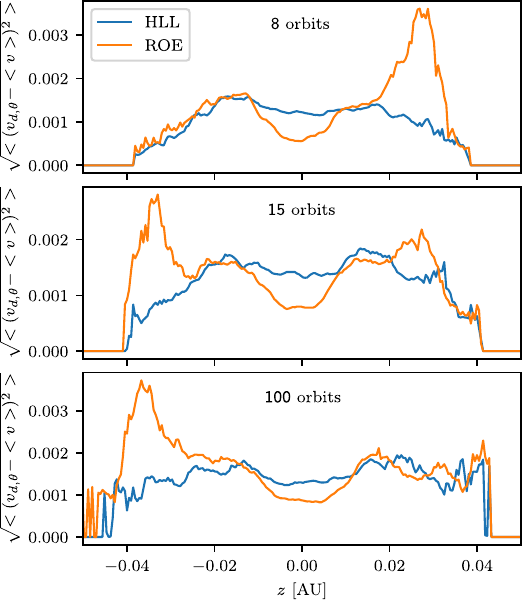  }
    \caption{RMS of the meridional dust velocity, $v_{\rm d, \theta}$. 
    The quantity $<v>$ corresponds to the mean of the meridional dust velocity to correct for velocity distributions that are not centered around 0. 
    To compute the RMS value at a given height, we consider all particles (and their velocities) which are located in a band at this height $\pm$ half the vertical size of grid cells.}
    \label{fig: RiemannSolversVelocityRMS}
\end{figure}

To understand where the additional midplane component for more accurate gas Riemann solvers (and in fact also higher order reconstruction schemes) comes from, we show the root-mean-square (RMS) of mean corrected meridional dust velocity as a function of height in Fig. \ref{fig: RiemannSolversVelocityRMS}. 
This RMS value quantifies the width of the velocity distribution.
Each panel represents a different stage in the disk's evolution: The first and second panels show the disk state where dust settling has stopped approximately, and the maximum dust-to-gas ratio in the midplane is reached for HLL ($\sim$8 orbits) and ROE ($\sim$15 orbits), respectively (see also bottom left panel in Fig. \ref{fig: 3RiemannSolvers}. 
The last panel shows a disk state in which turbulence has saturated for both solvers, and the midplane dust-to-gas ratio has reached equilibrium. 
Around the midplane and from 8 orbits onward, ROE leads to smaller RMS values and therefore less frequent high-velocity values, which allow the dust to settle more strongly. 
On the other hand, a wider velocity distribution for HLL indicates dust with higher velocities, which makes dust settling less efficient and likely prevents the formation of an additional midplane component in the dust density profile. 
At a height of approximately 0.02 AU, the width of velocity distributions of the two Riemann solvers are similar. 
The strong peak in RMS values for ROE, which appears only on one side, indicates broad velocity distributions that influence disk locations at heights of 0.03 AU and above. 
These are locations where dust density is strongly reduced relative to the midplane. 
The origin of these peaks remains an open question for future work.
 
In the 2D dust density plot for ROE, individual clumps in the midplane are already visible. 
Even though the outermost region for HLL and HLLC is not visible, we confirm that denser clumps are also forming at 10.6 AU for HLLC but not for HLL. 
The fact that these even denser clumps form only close to the outer boundary could indicate that they form due to a boundary effect. 
We discuss this issue in Sect. \ref{sec: outer_buffer_zone}.
We also refer to Sect. \ref{sec:midplane_dust_density_component} where we discuss the numerical origin of a midplane density component for HLLC and ROE, and the similar effect of a higher-order spatial reconstruction and more accurate Riemann solvers.

\subsection{Effect of the vertical domain size}
\label{sec: influenceDomainSize}
\begin{figure}[ht]
    \centering
    \includegraphics[]{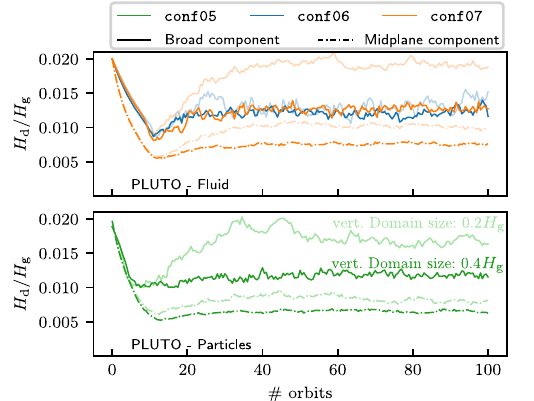}
    \caption{Test of the effect of the vertical domain size on the dust scale height for the PLUTO fluid (\textit{top}) and particle (\textit{bottom}) models. 
    The light colors (light green, light blue, and light orange) show the dust scale height for the original vertical domain size of 0.2 $H_{\rm g}$ and resolution of 4096$\times$512, as used in \cite{Flock2021}. 
    The dark colors (dark green, dark blue, and dark orange) correspond to the values for twice the vertical domain size and resolution of 4096$\times$1024 as used in this work. 
    Since the setups \texttt{conf05} and \texttt{conf07} exhibit a midplane density component, we show the dust scale height of this component as a dash-dotted profile.}
    \label{fig: influenceDomainSizePLUTOFluidParticles}
\end{figure}

As the total vertical domain size in \cite{Flock2021} is relatively small (0.2 $H_{\rm g}$), we investigate how the vertical domain size, in combination with the vertical boundary conditions, affects the dust scale height in PLUTO dust fluid and particle simulations.
The applied boundary conditions allow dust and gas (in dust-fluid simulations) and only gas (in particle simulations) to leave the domain in the vertical direction, but prevent their inflow. 
These boundary conditions have already been studied and discussed in multiple works; see, for instance, \cite{LiYoudinSimon2018}, \cite{Li2021} (their Sect. 4.1.4), and \cite{LimSimon20252D} (Appendix A), among others.
In Fig. \ref{fig: influenceDomainSizePLUTOFluidParticles}, we show the dust scale height for two different numerical setups for a PLUTO fluid (\texttt{conf06} and \texttt{conf07}) as well as one setup for particles (\texttt{conf05}). 
Each of these setups used a vertical domain of 0.4 $H_{\rm g}$ (shown in dark colors) and is compared to a simulation that used half the vertical domain size while keeping the resolution constant (faint colors).

From the results in Fig. \ref{fig: influenceDomainSizePLUTOFluidParticles}, we can see that a larger vertical domain (shown in dark colors) reduces the dust scale height significantly for fluid and particle simulations. 
This includes the dust scale height of the broad dust density component (solid line) and the midplane component (dash-dotted line). 
For the fluid simulation \texttt{conf06} in PLUTO, however, a larger vertical domain has a much smaller influence, still decreasing the dust scale height to a lesser extent.
These results agree well with \cite{Li2021}, who also showed that a larger vertical domain size leads to smaller dust scale heights (see their Sect. 4.1.4). 
In their work, the vertical domain of their standard box is also 0.4$\,H_{\rm g}$. 
Other works, such as \cite{YangJohansen2014} and \cite{LiYoudinSimon2018}, further demonstrate the importance of the vertical domain size. 
We briefly discuss this in Sect. \ref{sec:influence_vertical_domain_size}. 

\subsection{A comparison between PLUTO fluid and PLUTO particles}
\label{sec: pluto_fluid_pluto_particles}
In this section we investigate the differences between the two fundamental dust approaches (particles - \texttt{conf05} and fluid - \texttt{conf07}) when combined with more accurate numerical schemes.
The two dust approaches were simulated at increasing resolutions for the gas ($N_{\rm r}\times N_{\theta}$ = $512\times128$ to $8192\times2048$) and for vertical domain sizes of 0.4 $H_{\rm g}$. 
For the dust fluid, we used the same resolution as for the gas. 
\begin{figure*}
    \centering
    \includegraphics[]{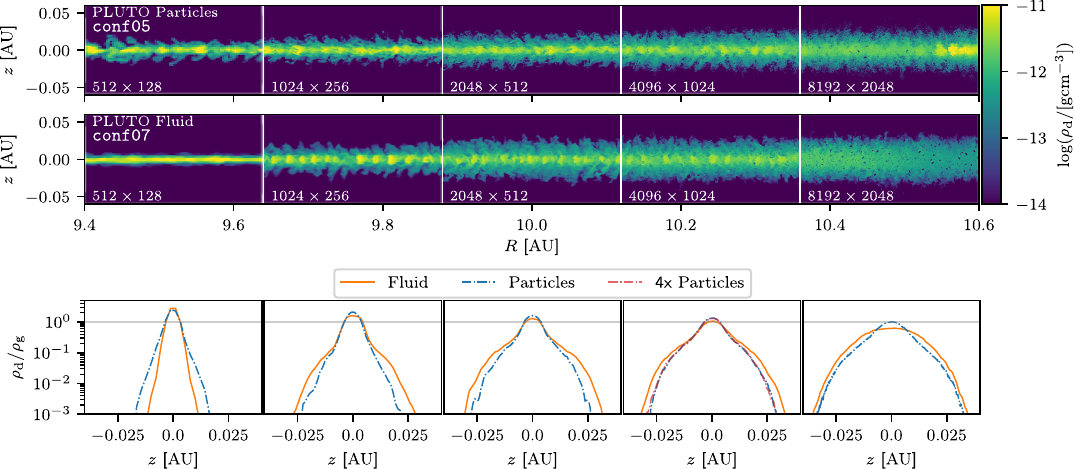}
    \caption{\textit{Top}: Comparison of the 2D dust density structure due to the nonlinear streaming instability for the particle approach (first row) and fluid approach (second row) in PLUTO. 
    For all resolutions, the disk is shown after 50 orbits (at 10 AU). 
    The vertical white lines indicate the radial positions at which we switch to another simulation for plotting. 
    The resolution of each simulation is indicated in the lower left corner of each section. 
    \textit{Bottom}: The vertical dust-to-gas ratio profiles for fluid simulations (solid line) and particle simulations (dash-dotted line), averaged in the radial domain 10 $\pm$ 0.15 AU. 
    The red dash-dotted line in the dust-to-gas ratio plot at a resolution of $4096\times1024$ shows the vertical profile from a simulation with 4 times the number of particles. 
    The horizontal gray lines at the bottom subplots indicate a dust-to-gas ratio of unity.}
    \label{fig: dustDensity2D_ComparisonFluidParticle_Roe}
\end{figure*}

In Fig. \ref{fig: dustDensity2D_ComparisonFluidParticle_Roe}, we show the 2D dust density structures for the fluid and particle approach for increasing resolution from left to right and after 50 orbits (determined at 10 AU). 
The last row shows the vertical dust-to-gas ratio for each resolution in both approaches. 
From the 2D density structures, it is already evident that the two fundamental approaches yield similar results, especially toward higher resolutions. 
Except for the highest resolution, they show individual clumps in the midplane, which contrast strongly with the background dust disk. 
The overall tendency to find a larger dust scale height with increasing resolution is evident in our dust-fluid and particle simulations and has already been reported in recent works \cite{LimSimon20253D} (see their Sect. 4.2) and \cite{LimSimon20252D} (see their Sect. 3.4).
However, the 2D density structures and the vertical dust-to-gas ratios indicate that except at the lowest resolution, particle simulations yield thinner dust disks with a smaller dust scale height for the broad component.

When examining the vertical dust-to-gas ratio profiles shown in the last row, which closely trace the vertical dust density profile because the gas density is nearly constant near the midplane, we observe that the midplane dust density component is clearly visible in both approaches.
This midplane component becomes broader with increasing resolution, and the dust-to-gas ratio decreases as the dust is distributed over a larger vertical domain. 
For the highest resolution of $8192\times2048$, corresponding to a resolution of $5120/H_{\rm g}$, the midplane component vanishes completely for the fluid approach, and the density structure indicates many small dust-depleted regions. 
These voids were also reported in \cite{Yang2017} for simulations with dust particles and a resolution of 2560/$H_{\rm g}$ (see their Fig. 1 first row). 
In our highest-resolution particle simulation, these void-like regions are just beginning to appear.

We also see that at heights between 0.02 and 0.025 AU, the dust-to-gas ratio decreases more strongly with height. 
As we investigate in Appendix \ref{sec: particles_per_cell}, this is linked to an average number of particles per cell of around 10 and smaller.

\begin{figure}
    \centering
    \includegraphics[]{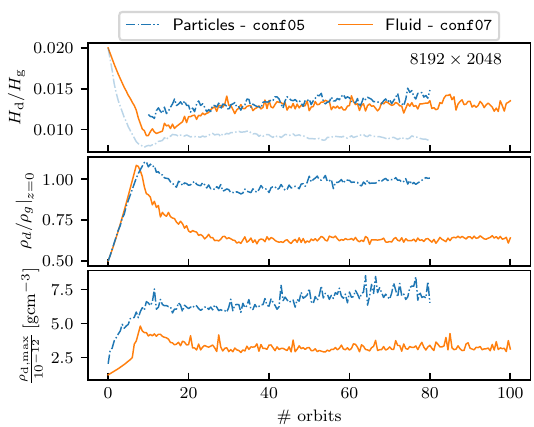}
    \caption{Comparison between the particle simulation ($\texttt{conf05}$, shown as dash-dotted lines) and fluid simulation ($\texttt{conf07}$, shown as solid lines) with a more accurate setup and a resolution of $8192\times2048$. 
    \textit{Top}: The dust scale heights as a function of orbits. 
    Dark blue shows the dust scale height for the full vertical range, while light blue corresponds to the midplane component. 
    \textit{Middle}: The evolution of the midplane dust-to-gas ratio. The dust-to-gas ratio was determined by radially averaging over the domain 10 $\pm$ 0.15 AU. 
    \textit{Bottom}: The evolution of the maximum dust density. The maximum dust density was determined in the radial domain between 9.4 and 10.4 AU.}
    \label{fig: dustScaleHeight_midDTGRatio_maxRhodust_FluidParticle_Roe}
\end{figure}

We will quantitatively investigate the temporal evolution of the dust scale height between the two approaches as shown in the upper plot of Fig. \ref{fig: dustScaleHeight_midDTGRatio_maxRhodust_FluidParticle_Roe} for the highest resolution. 
We show the dust scale height of the entire dust disk (in dark blue) and that of the midplane component (in light blue). 
The particle simulation yields a slightly larger dust scale height for the broad component; however, the difference compared to the fluid simulations is small ($H_{\rm d} / H_{\rm g} = 0.0134$ for particles versus $H_{\rm d} / H_{\rm g} = 0.0129$ for the fluid over the full vertical domain). 
Because the midplane component for the fluid simulation no longer exists, we do not show the temporal evolution here.
We refer to Sect. \ref{sec: convergence} for the convergence check between the two approaches. This check was made for both density components individually.

In addition to the dust scale height, we also compare, in Fig. \ref{fig: dustScaleHeight_midDTGRatio_maxRhodust_FluidParticle_Roe}, the midplane dust-to-gas ratio (shown in the middle panel) and the maximum dust density (shown in the bottom panel) for particle (dash-dotted line) and fluid (continuous line) simulations for the highest resolution case. 
Comparing the midplane dust-to-gas ratio, we find that once the instability reaches saturation (approximately 30 orbits), the particle simulations systematically yield higher values than the fluid simulations. 
While with particles, we find even values above unity, the fluid simulation stays between 0.5 and 0.75. 
This is linked to the particle simulation, still showing a midplane component in the density, resulting in a higher dust concentration at this location.

Examining the evolution of the maximum dust densities in the bottom panel of Fig. \ref{fig: dustScaleHeight_midDTGRatio_maxRhodust_FluidParticle_Roe}, we find that the particle approach generally yields higher peak densities, which can temporarily exceed $7.5\cdot10^{-12}\ \mathrm{g cm^{-3}}$. 
For a dust fluid, we find values that stay around $3\cdot10^{-12}\ \mathrm{g cm^{-3}}$.
In addition, the particle profile of maximum dust density continues to increase with the simulated number of orbits.
These maximum densities are most likely located in the denser clumps, which form more easily in dusty regions where dust disks are thinner, and the higher dust-to-gas ratio facilitates clump formation. 
Once a saturated state of the instability is reached (from 30 orbits onward), the temporal average of the maximum dust density is approximately 50$\%$ of the value for the particle approach.
In the highest-resolution 2D density snapshot of the particle simulations shown in Fig. \ref{fig: dustDensity2D_ComparisonFluidParticle_Roe}, dense clumps are visible slightly beyond 10.5 AU. 
These clumps are not included in the analysis of the maximum dust density and are likely to have formed at the outer boundary before migrating inward.
Such clumps are not visible in the highest-resolution dust-fluid case, where the midplane density decreases toward the outer radial boundary. 
The likely reason is that insufficient material was refilled at this boundary, which drifts inward. 
We refer to Sect. \ref{sec: outer_buffer_zone}, where we discuss dust replenishment at the outer boundary.

\cite{Baronett+2026Pfits} studied the unstratified shearing box BA test from \cite{JohansenYoudin2007} for a highest resolution of $1024^{2}$. When they computed the maximum dust density separately in their fluid and particle simulations, they found only a 30$\%$ lower value in the fluid simulations. 
This is a smaller discrepancy than for values around 50$\%$. We come back to this in Sect. \ref{sec: comparison_to_other_work}.

\begin{figure}[]
    \centering
    \includegraphics[]{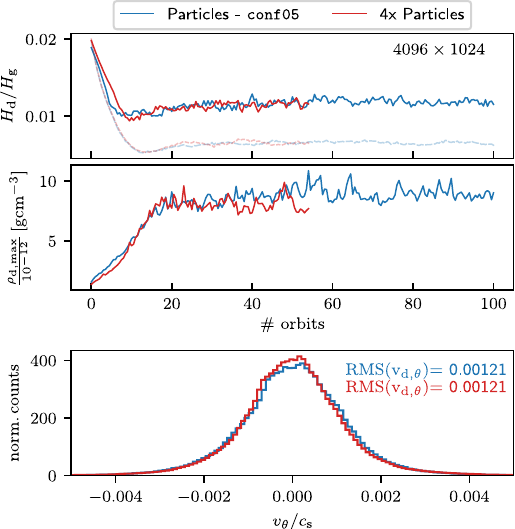}
    \caption{Comparison of dust scale height (\textit{top}), maximum dust concentration (\textit{middle}), and meridional dust velocity distribution (\textit{bottom}) for setup $\texttt{conf05}$ with 1.04$\cdot10^{7}$ particles (blue lines) and with 4.16$\cdot10^{7}$ particles (red lines). 
    In the first panel, dark, solid lines indicate the dust scale height of the broad density component, while light, dashed lines correspond to the midplane density component. 
    The distributions in the bottom panel consider velocity data only within [9.4, 10.4] AU in the radial direction and are taken after 50 orbits of evolution.}
    \label{fig: 3quantities_1x_4x}
\end{figure}

In the bottom panel of Fig. \ref{fig: dustDensity2D_ComparisonFluidParticle_Roe}, for a resolution of $4096\times1024$ and 1.04$\times 10^{7}$ particles, we have already seen that four times more particles hardly change the vertical profile of the dust-to-gas ratio. 
In Fig. \ref{fig: 3quantities_1x_4x} we now compare the measured dust scale height, maximum dust concentration, and meridional dust velocity distribution. 
The results show that an increase in initial particle number, so that we start with 9.9 particles per cell, does not change characteristic properties of the dust disk.

\subsection{A comparison between PLUTO fluid and FARGO3D}
\label{sec: pluto_fluid_fargo3d}
\begin{figure*}[]
    \centering
    \includegraphics[]{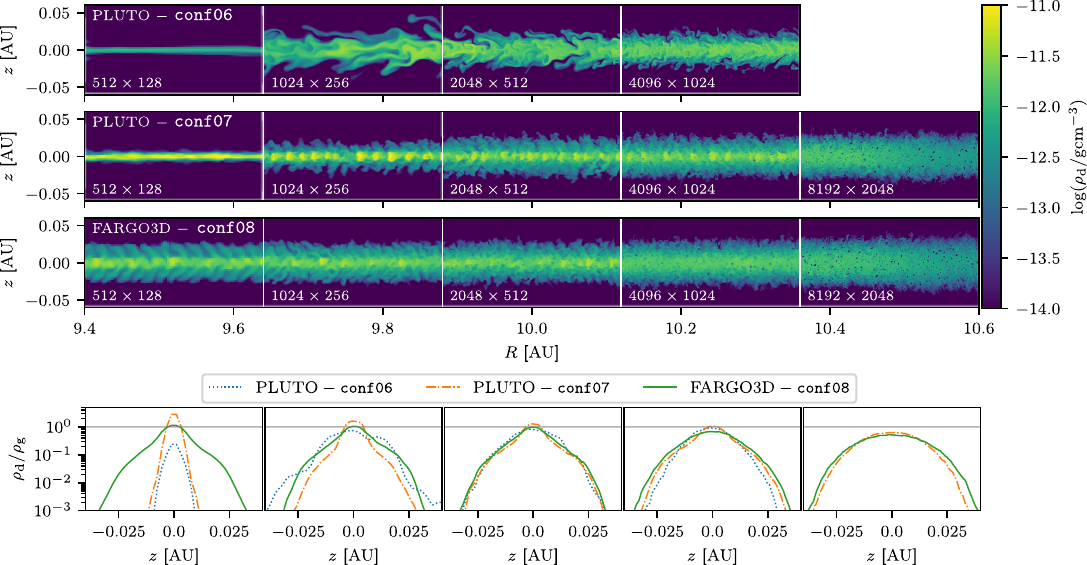}
    \caption{\textit{Top three rows}: 2D dust densities for dust fluid simulations and increasing resolution (information shown on the left side of each panel). 
    The vertical white lines indicate the transition to another simulation. 
    \textit{Bottom}: The vertical dust-to-gas ratio profiles for each simulation, averaged over the radial domain 10 $\pm$ 0.15 AU. 
    The solid line represents FARGO3D (\texttt{conf08}) results, the dashed-dotted line represents PLUTO fluid (\texttt{conf07}) results, and the dotted line represents PLUTO fluid (\texttt{conf06}) results. 
    The horizontal gray line indicates a dust-to-gas ratio of unity.  
    The disks are shown after 50 orbits (at 10 AU).}
    \label{fig: dustDensity2D_FARGO3D}
\end{figure*}

Because PLUTO and FARGO3D allowed us to model and evolve dust as a fluid, we were able to verify whether they yielded similar results for the dust scale height, maximum dust density, and midplane dust-to-gas ratio. 
The PLUTO setups used for comparison treated dust as a fluid using a less (\texttt{conf06}) and more (\texttt{conf07}) accurate configuration, where the latter is already shown in the top panel of Fig. \ref{fig: dustDensity2D_ComparisonFluidParticle_Roe}.
Additionally, for the same increasing resolutions, we performed simulations using FARGO3D (\texttt{conf08}). 
For this hydrodynamics code, the simulation for a resolution of 4096$\times$1024 is the same as shown in the top panel of Fig. 1 in \cite{Ostertag2025StrongClumping}. 
The results for the 2D dust densities and the vertical dust-to-gas ratio profiles are shown in Fig. \ref{fig: dustDensity2D_FARGO3D}.

When first looking at the 2D density results for the least accurate configuration of PLUTO (\texttt{conf06}, top panel in Fig. \ref{fig: dustDensity2D_FARGO3D}), we can see that the structure differs significantly from the results of the more accurate PLUTO fluid configuration (\texttt{conf07}, second row in the same figure) or the results with FARGO3D (\texttt{conf08}, third row). 
For the two lowest-resolution cases, the \texttt{conf06} setup yields substantially lower dust densities overall, which is also reflected in the last row of Fig. \ref{fig: dustDensity2D_FARGO3D}, showing the vertical dust-to-gas ratio profile after 50 orbits: While more accurate PLUTO fluid and FARGO3D setups have dust-to-gas ratio values larger or roughly equal to 1, \texttt{conf06} reaches values around 0.2. 

At higher resolutions, the less accurate PLUTO setup \texttt{conf06} does not show the additional midplane component in the dust density. 
This is no surprise, as we have seen already in Fig. \ref{fig: spatialReconstruction} and \ref{fig: 3RiemannSolvers} that linear spatial reconstruction and the HLL solver do not show this midplane component individually for dust particles, and the combination of the two does not give us a different structure, even for a dust fluid. 

Inspection of FARGO3D simulation results shows that the additional midplane component in the dust density is visible at the three lower resolutions ($512\times128$, $1024\times256$, and $2048\times512$), as indicated also by an additional Gaussian component around the midplane in the vertical dust-to-gas ratio profile (lowest row in Fig. \ref{fig: dustDensity2D_FARGO3D}). 
However, for resolutions of 4096$\times$1024 or higher, this component vanishes. 
For both resolutions, we also observe void-like regions, as in \cite{Yang2017} (see the first row of their Fig. 1 for a similar set of parameter values for resolution, global dust-to-gas ratio, and Stokes number as in our work). 
We refer to Appendix \ref{sec: depleted voids} for the lifetime and trajectories of these voids. 
Overall, the results show that the dust scale height of the broad density component increases only slightly, from 0.0127$H_{\rm g}$ at the lowest resolution to 0.0141$H_{\rm g}$ at the highest resolution.

When we compare the results between the two PLUTO fluid configurations and FARGO3D, we see that the density structure of FARGO3D and PLUTO \texttt{conf07} are more similar than compared to the least accurate PLUTO setup (\texttt{conf06}). 
At the highest resolution, the density structure is almost identical, with a slightly thinner disk in PLUTO, having a dust scale height of 0.0129$H_{\rm g}$. 
The similarity between the two codes vanishes at lower resolutions. 
For $512\times128$ cells, FARGO3D shows a significantly thicker dust scale height of 0.0127$H_{\rm g}$ compared to 0.0048$H_{\rm g}$ for PLUTO. 
For this resolution, the number of grid cells covering the broad density component is 4 for FARGO3D and 2 for PLUTO.

Finally, we observe that in the highest-resolution FARGO3D cases and in the most accurate PLUTO setup, \texttt{conf07} at 50 orbits, insufficient material is delivered from the outer radial domain. 
This results in a reduced dust density in the midplane close to the outer edge of the domain.
\begin{figure}[ht]
    \centering
    \includegraphics[]{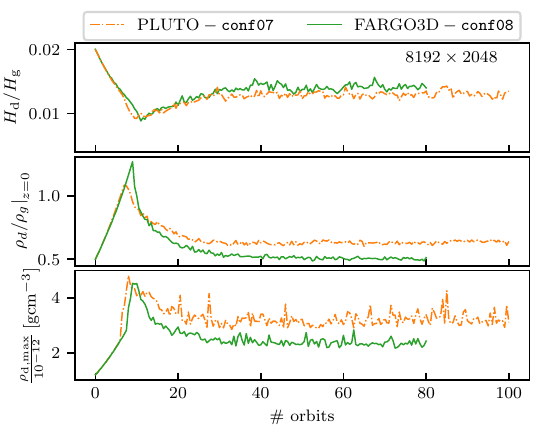}
    \caption{Comparison between FARGO3D and PLUTO simulations at the highest resolution of $8192\times2048$. 
    \textit{Top}: The temporal evolution of the dust scale height $H_{\mathrm{d}}$ in units of gas scale height $H_{\mathrm{g}}$. 
    \textit{Middle}: The temporal evolution of the average midplane dust-to-gas ratio in the radial domain 10 $\pm$ 0.15 AU. 
    \textit{Bottom}: The temporal evolution of the maximum dust density in the radial domain 9.4 to 10.4 AU. 
    The solid line represents the FARGO3D \texttt{conf08} simulation, while the dashed-dotted lines represent the simulations for PLUTO \texttt{conf07}.}
    \label{fig: QuantitativeComparisonDustScaleHeight_dtg_maxdens}
\end{figure}

Similar to Fig. \ref{fig: dustScaleHeight_midDTGRatio_maxRhodust_FluidParticle_Roe}, we compare the vertical dust density scale height, the midplane dust-to-gas ratio, and the maximum dust density between the PLUTO fluid (\texttt{conf07}) and the FARGO3D setup for a resolution of $8192\times2048$ in Fig. \ref{fig: QuantitativeComparisonDustScaleHeight_dtg_maxdens}. 
As shown in the 2D density structure, neither setup has an additional Gaussian component in the midplane. Hence, we plot only the temporal evolution of the broad density component in the first panel. 
Both simulation runs (\texttt{conf07} and \texttt{conf08}) show similar dust scale heights within 30 orbits. 
However, when a saturated state was reached (roughly at 30 orbits), the dust scale height of the broad density component using FARGO3D is slightly larger, with a temporal average of 0.0141 $H_{\rm g}$ compared to 0.0129 $H_{\rm g}$ for PLUTO. 
These scale heights are covered by 72 and 66 cells for FARGO3D and PLUTO, respectively.

In the middle and bottom panels, we compare the dust-to-gas ratio in the midplane and the maximum dust density, respectively. 
At the highest resolution, the most accurate PLUTO (\texttt{conf07}) setup still yields larger values for the dust-to-gas ratio and the maximum dust density. 
This is linked to the result that setup \texttt{conf07} has a thinner dust disk, leading to a slight dust enhancement in the midplane.
Despite yielding different results, the maximum dust density (obtained with PLUTO's most accurate setup) is only 32$\%$ higher than the maximum dust density of FARGO3D after 50 orbits.

\begin{figure*}[ht]
    \centering
    \includegraphics[]{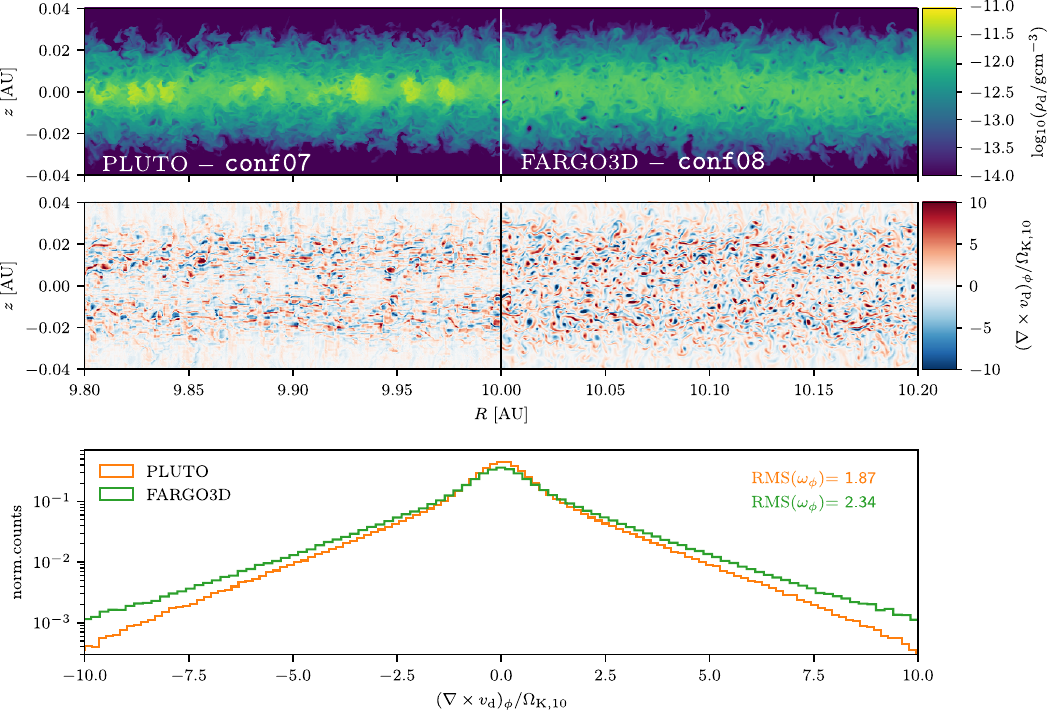}
    \caption{Comparison of the vorticity between the PLUTO fluid (\textit{left}) and FARGO3D (\textit{right}) simulation at resolution of $4096\times1024$. 
    \textit{Top}: The 2D dust density structure. \textit{Middle}: The 2D dust vorticity in the azimuthal direction. 
    The vertical white and black lines at 10 AU separate the results of the two codes. 
    \textit{Bottom}: The distributions of the dust vorticity. 
    The RMS values indicate the RMS of the vorticity distribution. 
    For all panels, the data were taken at 50 orbits. 
    The vorticity is normalized with $\Omega_{\rm K, 10}$, the Keplerian angular velocity at 10 AU.}
    \label{fig: vorticity_4096x1024}
\end{figure*}

As a final comparison between FARGO3D (a ZEUS-like code) and PLUTO, we look at the difference in the dust vorticity, as both simulations indicate the existence of dust-depleted voids or vortices. 
To compute the vorticity, we used a second-order-accurate central-difference scheme that accounts for spherical geometry. 
Fig. \ref{fig: vorticity_4096x1024} shows the 2D dust density structure for both codes (top panel), the azimuthal component of the 2D dust vorticity (middle panel), and a comparison of the normalized distributions (bottom panel). 
The result shows that an existing midplane component in the dust density likely reduces the vorticity in the midplane, causing smaller values to appear more frequently. 
In addition, the bottom panel shows that FARGO3D has a slightly higher occurrence rate of higher vorticity values. 

\subsection{Convergence Test}
\label{sec: convergence}
In this section we explore how the different codes and dust approaches influence disk properties like the dust scale height, the meridional dust velocity, and the RMS of vorticity. 
These properties will be shown at increasing resolution to determine whether the quantities converge to a particular state. 
The relevant setups we consider here are PLUTO Particles (\texttt{conf05}), PLUTO Fluid (\texttt{conf07}), and FARGO3D (\texttt{conf08})

\begin{figure*}[ht]
    \centering
    \includegraphics[]{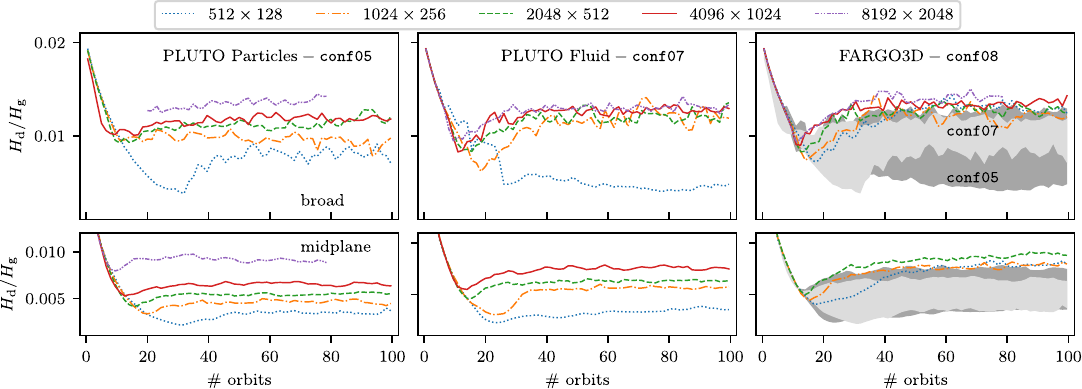}
    \caption{Evolution of the dust scale height for three different numerical setups, as indicated in the upper center of every panel in the first row, and for increasing resolution. 
    \textit{Top}: The evolution of the dust scale height of the broad density component. 
    \textit{Bottom}: The evolution of the dust scale height of the midplane component. 
    The shaded areas in the subplots for FARGO3D indicate the domain regions covered by PLUTO particle (light gray) and PLUTO fluid (dark gray) simulations for a resolution of $4096\times1024$ to facilitate comparison across all three setups. 
    For FARGO3D and PLUTO Fluid, we do not show the dust scale height of the midplane component at the two highest and highest resolutions, respectively, as this component has disappeared.}
    \label{fig: Convergence}
\end{figure*}

To this point, we have either qualitatively examined 2D dust densities from different approaches and codes with increasing resolution or quantitatively compared measured dust scale heights across approaches and codes at one fixed resolution.  
In Fig. \ref{fig: Convergence}, we quantitatively investigate whether any approach or code exhibits convergence in the dust scale height as the resolution is increased. 
Each setup is shown in a separate column. 
The Top row shows the evolution of the dust scale height for the broad density component, whereas the bottom row shows the evolution of the midplane component. 
To better compare the three setups, we highlight, in the last column, the spanned area of \texttt{conf05} and \texttt{conf07} in light and dark gray, respectively. 
This only considers the results up to a resolution of $4096\times1024$. 

From the values of the broad dust density component, we can generally see that the dust scale height evolves differently with resolution between the three setups. 
In both dust approaches in PLUTO, increasing resolution increases the scale height, whereas in FARGO3D, at the lowest resolution, the scale height is already close to that at the highest resolution. 

A comparison between FARGO3D and PLUTO fluid shows that at a resolution of $2048\times512$, the two setups indicate a convergence. 
This is different from the results for PLUTO particles, where we consistently observe an increase in the scale height, even at resolutions of $4096\times1024$ and $8192\times2048$. 
However, it should be noted that the dust scale height at the highest resolution differs little across all setups (given the by-eye variation in the temporal evolution), indicating convergence toward a general finite value. 
For PLUTO particles, PLUTO fluids, and FARGO3D, the temporal averaged value is 0.0134 $H_{\rm g}$, 0.0129 $H_{\rm g}$, and 0.0141 $H_{\rm g}$, respectively.

However, we observe that the dust scale height of the midplane component continues to increase with increasing resolution across all three setups, with no evidence of convergence. 
For FARGO3D, the midplane component already disappears completely at a resolution of $4096\times1024$, whereas for the PLUTO fluid, this occurs only at the highest resolution. 
Therefore, in the cases where the midplane component vanishes, we do not show its temporal evolution. 
If the dust scale height is computed using Eq. \ref{eqn: ComputeDustScaleHeight} - thereby combining the broad and midplane components - we expect an increasing value with resolution for all three setups.

\begin{figure}[ht]
    \centering
    \includegraphics[]{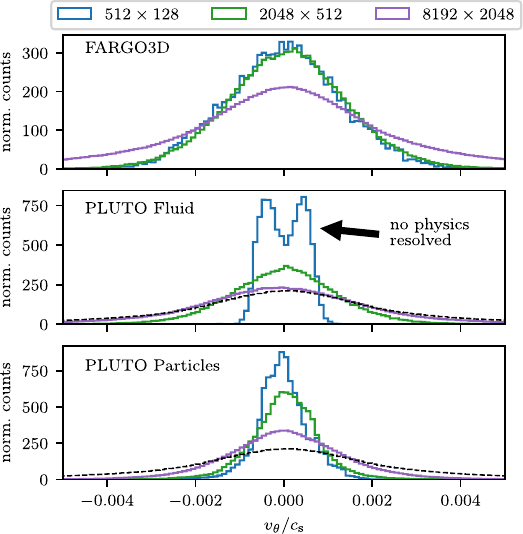}
    \caption{Distribution of meridional dust velocity for FARGO3D (\textit{top}), PLUTO fluid, and particle simulations (\textit{middle} and \textit{bottom}, respectively) for three resolutions. 
    The dashed black line in the two PLUTO panels represents the distribution at the highest FARGO3D resolution to enable a better comparison. 
    All distributions are shown after 50 orbits of disk evolution and only considered data in the radial domain between 9.4 and 10.4 AU and in the vertical domain between 0.05 AU above and below the midplane. 
    The velocity is given in units of sound speed at 10 AU.}
    \label{fig: Convergence of dust velocity}
\end{figure}

From Fig. \ref{fig: Convergence} and Fig. \ref{fig: dustDensity2D_FARGO3D}, we saw that the dust disk for FARGO3D at the lowest resolution is much thicker compared to PLUTO fluid and particle simulations, and the dust scale height of the broad density component is already close to the value we find for the highest resolution. 
To identify why FARGO3D seems to converge faster than PLUTO, we look first at the meridional dust velocity, which is closely connected to dust scale height (a higher bulk velocity should allow dust to reach higher altitudes). 
Figure \ref{fig: Convergence of dust velocity} shows the corresponding distributions for three different resolutions (going from lowest to mid to highest resolution) for FARGO3D (top panel), PLUTO fluid (middle panel), and PLUTO particle simulations (lowest panel). 
For a better comparison between FARGO3D and PLUTO, we plot the distribution of FARGO3D at the highest resolution in the two PLUTO panels as a dashed black line. 
The results show two different aspects. 
Firstly, all three simulations show different distributions at the lowest resolution. 
However, as the resolution increases, they approach the final distribution of FARGO3D. 
Secondly, the distribution of FARGO3D at the lowest resolution is closer to the final distribution compared to that of PLUTO. 
In general, an increase in resolution makes higher velocity values more abundant, potentially allowing the dust to reach higher altitudes.

\begin{figure}[ht]
    \centering
    \includegraphics[]{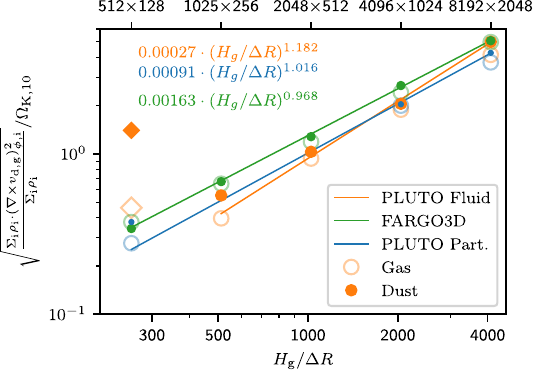}
    \caption{Comparison of the density-weighted RMS of the vorticity distributions. 
    The results for gas and dust are shown with open and filled circles, respectively. 
    For the PLUTO fluid simulation at the lowest resolution, we show the data using open and filled squares, since no physics is resolved. 
    The squared data points were not used to fit a power law. 
    The quantity $v_{\rm d,g}$ denotes the corresponding species velocity (dust or gas) used to compute either the dust or gas vorticity. 
    The variable $\rho_{i}$ is the corresponding species volume density at grid cell $i$. 
    The vorticity was normalized with $\Omega_{\rm K, 10}$, the Keplerian angular velocity at 10 AU. 
    The data were taken after 50 orbits of evolution, and we considered only vorticity values in the radial direction within [9.4, 10.4] AU and in the vertical direction within [-0.05, 0.05] AU.}
    \label{fig: convergence rms vorticity}
\end{figure}

In addition to looking at the bulk velocity as a way to produce a higher dust scale height, we investigate the level of vorticity due to turbulence. 
In Fig. \ref{fig: convergence rms vorticity} we show the density-weighted RMS of the dust and gas vorticity distribution as a function of $H_{\rm g} / \Delta R$, the number of radial grid cells per gas scale height. 
The density weighting allows us to focus on dust-enriched regions. 
Due to partial data loss, we plot PLUTO particle results only for the resolutions $512\times128$, $4096\times1024$ (here the simulation with properly rescaled number of particles is used), and $8192\times2048$. 
Because we do not resolve the instability at the lowest PLUTO fluid resolution, we highlight this value with a different marker. We also do not use this lowest-resolution value to determine a power law for this setup. 
The results indicate that the vorticity distributions of PLUTO and FARGO3D broaden as resolution increases. 
For resolutions between $512\times128$ and $8192\times2048$, FARGO3D yields larger values, but at the highest resolution, PLUTO fluid and FARGO3D results are almost identical. 
For all three setups, the power-law trend of the form $a \, x^{b}$ is clear. 
The values of $a$ and $b$ for each setup are shown in the figure itself. As the data follow power laws, there is no sign of vorticity saturation.

The results of Figs. \ref{fig: Convergence of dust velocity}, \ref{fig: convergence rms vorticity}, and \ref{fig: vorticity velocity lowest res} indicate that dust vorticity is not the reason for the initially very different dust scale heights between PLUTO and FARGO3D, as, for the lowest resolution, their RMS values are similar. 
Instead, the distribution of meridional dust velocity is much broader for FARGO3D with RMS values almost twice as large as for PLUTO particle simulations. 
At lower resolutions, these higher bulk velocities allow the dust to reach greater altitudes, leading to a larger scale height.

\section{Discussion}
\label{sec: Discussion}

\subsection{Midplane dust density component}
\label{sec:midplane_dust_density_component}
The purpose of Sects. \ref{sec: reconstruction_schemes} and \ref{sec: riemann_solver} was to evaluate the impact of numerical accuracy on the outcome of density structures. 
More specifically, we investigated the influence of the spatial reconstruction scheme and different Riemann solvers in PLUTO particle simulations. 
A higher-order reconstruction scheme and a more accurate gas Riemann solver both allow the dust to settle more strongly to the midplane, showing an additional midplane dust density component for intermediate resolutions. 
For Riemann solvers, this result can be counterintuitive: A more diffusive solver should lead to lower gas velocities, which are not able to mix the dust efficiently in the vertical direction. 
With Fig. \ref{fig: RiemannSolversVelocityRMS}, we have shown that in the midplane a more diffusive solver, like HLL, systematically leads to a broader velocity distribution with larger velocities, compared to a more accurate solver. 
However, the numerical origin is not yet clear. The difference between HLL and HLLC / ROE is that the latter two do preserve discontinuities of momentum in the transverse direction to the grid cell surface, which are smeared out with HLL. 
This smearing has a direct influence on the gas vorticity distribution (see bottom panel in Fig. \ref{fig: VorticityRiemannSolver_SpatialRecon}), as HLL leads to more frequent values around zero and therefore weaker vortices compared to ROE.
We find a similar result between PLM and PPM: a more accurate scheme yields higher vorticity values and narrower velocity distributions, such that the dust density shows an additional midplane component.
Whether and how vorticity explains the difference in dust settlement is also unclear, and future research on this aspect is necessary.
In the midplane dust-to-gas ratio panels in Fig. \ref{fig: spatialReconstruction} and \ref{fig: 3RiemannSolvers}, we have noticed a shorter dust settling time for a less accurate spatial reconstruction scheme and Riemann solver before turbulence is able to puff the disk up. The influence of this aspect on the presence of an additional midplane dust density component remains unclear and needs to be investigated in future work.

We also noted that the midplane dust-to-gas ratio for a higher-order spatial reconstruction scheme (PPM) is very similar to that for a more accurate Riemann solver (HLLC, ROE), with values just above 1. 
This raises the question of whether the lower truncation errors of a piecewise parabolic reconstruction scheme are sufficient to reduce the smearing of the discontinuity in transverse momentum (or numerical diffusivity) introduced by the HLL scheme. 
This momentum discontinuity, however, is conserved with HLLC. 
We could test whether similar results can be achieved when choosing the HLL solver in combination with a linear reconstruction scheme and an even higher resolution than in this work, compared to the HLLC solver with linear reconstruction and a moderate resolution. 

Even though HLLC is able to show the additional midplane component of the dust density, it also exhibits large diffusive errors when the gas flow is subsonic \citep[see Sect. 2.4.3 in][]{LeidiAndrassy2024}. 
This condition is fulfilled in our simulations, with typical Mach numbers below 1$\%$. 
For this reason, the previously mentioned work follows the approach from \cite{MinoshimaMiyoshi2021} and presents a modified version of HLLC, called low-dissipation HLLC or LHLLC, where the ratio of numerical diffusion to the amplitude of pressure fluctuations remains independent of the Mach number. 
At a low additional cost, this modified solver can improve the accuracy of low-Mach-number simulations. However, its performance in coupled gas-dust simulations needs to be tested in future work.

\subsection{Effect of the vertical domain size}
\label{sec:influence_vertical_domain_size}
The vertical domain size of 0.2$H_{\rm g}$, that we used in Sects. \ref{sec: reconstruction_schemes} and \ref{sec: riemann_solver}, is the same as used in \cite{Flock2021}. 
This size is smaller than the 0.4$H_{\rm g}$ used in \cite{Li2021}, who showed that smaller domain sizes produce larger dust scale heights. 
We have verified this for our disk setup in Fig. \ref{fig: influenceDomainSizePLUTOFluidParticles} for particle and fluid simulations using PLUTO. 
This means that the true dust layers for the different spatial reconstructions and Riemann solvers will be smaller, hence we expect to find even higher midplane dust-to-gas ratios and midplane dust densities. 
However, all setups (\texttt{conf01} - \texttt{conf05}) are affected in the same way, and any differences in the midplane dust-to-gas ratio or maximum dust density should arise only from different reconstruction schemes or Riemann solvers. 
Therefore, we should see these results only as a relative comparison.

To further emphasize the importance of a meaningful selection of the vertical domain size, we briefly discuss the results from \cite{LiYoudinSimon2018}. 
While clumping remains persistent across nearly all their vertical boundary conditions (VBCs) and box sizes, the particle scale height $H_{\rm p}$ varies widely across different VBCs and box sizes (see their Fig. 4). 
But within a specific time window they considered (called stage 3 and indicated by two cyan vertical lines in Fig. 4), $H_{\rm p}$ is similar across VBCs when $L_{\rm z}=0.8 H_{\rm g}$. 
This result shows that increasing $L_{\rm z}$ reduces artifacts from vertical boundary conditions. 
Since they were working with a Stokes number 30 times larger than in the current work, stronger dust settling decreases the fraction of the computational domain influenced by particles, effectively mimicking a larger vertical domain. 

\subsection{Effect of the outer buffer zone}
\label{sec: outer_buffer_zone}
The 2D density structures for FARGO3D and the PLUTO fluid approach also showed that insufficient dust was delivered from the outer buffer zone, leading to a drop in midplane dust density. 
This was not observed in particle simulations, which show even denser clumps forming at the outer edge of the domain. 
Even though both approaches used the same dust refilling technique in the outer buffer zone, as mentioned in \cite{Flock2021}, the results differ in their effects on the dust disk. 
In Appendix \ref{sec: dustDiskMass} we show that for particle simulations, we gain dust mass in time, while for fluid simulations (PLUTO and FARGO3D), we lose dust mass. 
To make a reliable comparison of quantities like midplane dust-to-gas ratio or maximum dust density, the radial domain in which boundary effects are visible must be disregarded. 
We did this by neglecting the simulation results outside 10.4 AU. 
In Fig. \ref{fig: dustDensity2D_FARGO3D} (first and last row), we saw that for the less accurate PLUTO setup $\texttt{conf06}$ and resolutions of $512\times128$ and $1024\times256$, a significant amount of dust was lost at 100 orbits, despite using the refilling in the outer buffer zone. 
This results in a weaker interaction via drag, therefore reducing the strength of resulting instabilities. 
Hence, the direct comparison between the three setups remains difficult. 
In future work, a more effective dust refilling mechanism needs to be implemented. One possibility is to fix the dust mass flux at the outer boundary, for example, as in \cite{SchoonenbergOrmel2017} or \cite{WangOrmel2025} and references therein, instead of refilling the dust in the damping zone.

\subsection{Effect of the temporal and spatial accuracy}
When we revisit the comparison between the more accurate PLUTO fluid and FARGO3D simulations, the question arises as to why the density structure of the two codes is similar. 
FARGO3D achieved first-order time accuracy \citep[as shown in Fig. 2 of][for a set of six fluids]{Benítez-Llambay2019} and second-order spatial accuracy. 
In contrast, the corresponding configurations of PLUTO were third- and second-order accurate in time for gas and dust \citep[for dust with explicit dust integrator see Appendix A1 in][]{Ziampras2025DustFluidModule}, respectively. 
Their spatial accuracy was of fourth order for gas due to the spatial reconstruction scheme and of second order for dust \citep[see Appendix A2.3 in][]{Ziampras2025DustFluidModule}. 
Mixtures of different orders for temporal and spatial accuracy will presumably average out, but they should be higher than for FARGO3D. 
This brings up numerical diffusivity for FARGO3D due to lower-order schemes and a low number of cells per dust scale height at the lowest resolution (128$/(0.4H_{\rm g}$) and a dust scale height of 0.0127 $H_{\rm g}$ results in roughly 4 cells per $H_{\rm d}$). 
However, in this case, we should also see numerical diffusivity dominating the disk structure in the less accurate setup of PLUTO ($\texttt{conf06}$), where we find a temporal average dust scale height of 0.0044 $H_{\rm g}$ (see Table \ref{tab: overview}). 
An alternative explanation is that FARGO3D better resolves the instability's turbulent structures. 
This may not be surprising as FARGO3D is a ZEUS-like code. 
Despite \cite{SeligmanLaughlin2017} arguing that PLUTO and ZEUS both suffer from severe vortex dissipation, their Fig. 3 shows that the vortex dissipation is lower for ZEUS than for PLUTO. 
In Fig. \ref{fig: vorticity_4096x1024} and Fig. \ref{fig: convergence rms vorticity}, we have shown that FARGO3D is indeed able to produce larger vorticity values compared to PLUTO, which supports their statement of vortex dissipation.
At the lowest resolution, the RMS vorticity value for PLUTO particles is comparable to that of FARGO3D, even though the dust disk for particles has a smaller scale height. 
This shows that vorticity does not explain the difference in scale height between the two codes at the lowest resolution. 
As we have shown in Fig. \ref{fig: Convergence of dust velocity}, the low-resolution distribution of the dust meridional velocity in FARGO3D is much closer to the high-resolution distribution, whereas for PLUTO fluid and particles, it initially deviates significantly. 
Why the velocity distributions for FARGO3D are already close to the final distribution remains an open question and needs to be identified with future work.

\subsection{Dust-depleted voids}
The PLUTO fluid and FARGO3D simulations indicate a convergence of the dust scale height for the broad density component, as well as a vanishing midplane component that is replaced by a field of many dust-depleted voids. 
In Appendix \ref{sec: depleted voids} we examine the dust-depleted voids in more detail. 
Tracking the trajectories of 8 voids by hand over 1 orbit, we find that some voids tend to migrate in a certain direction, while others stay close to the location where we started tracking. 
Performing a temporal average of the dust density over different time windows, we find that no void structure survives a time window of 2 orbits. 
While voids, moving in a fixed direction, do not contribute much to the average density field, other voids that tend to circularize around a fixed position will dissolve after around 2 orbits. 
Considering the duration of our simulations of about 100 orbits, these voids cannot be considered as persistent structures and undergo merging with other voids or dissolve.

\subsection{Effect of particle sampling}
Because we fixed the number of particles in our simulations despite increasing the resolution, the particle sampling per cell decreases. This should enhance the deviation from results with a dust fluid. 
In addition, as dust particles settle toward the midplane, the particle-resolved region shrinks, reducing the strength of the streaming instability and the induced turbulence in the upper layers.
In this way, dust particles cannot reach higher altitudes.

To test the impact of proper particle sampling, we created a new setup of $\texttt{conf05}$, using four times the number of particles. 
This yields an average of 9.9 particles per cell. 
However, as we have shown in Fig. \ref{fig: dustDensity2D_ComparisonFluidParticle_Roe} (see the bottom panel for a resolution of $4096\times1024$) and Fig. \ref{fig: 3quantities_1x_4x}, a higher number of particles does not significantly change characteristic properties such as dust scale height, maximum dust concentration, or velocity distribution. 
Even though we did not adapt the number of particles for the highest-resolution run, which would require evolving 16.6$\times 10^{7}$ particles, we do not expect to see a significant change in the dust disk properties. 

\subsection{Convergence of the dust scale height and vorticity distribution}
The convergence study of the dust scale height for particle methods in Fig. \ref{fig: Convergence} indicates that the broad density component at the highest resolution should be comparable to those of FARGO3D, for instance. 
However, Fig. \ref{fig: dustDensity2D_FARGO3D} and \ref{fig: dustDensity2D_ComparisonFluidParticle_Roe} show that the disk is smaller in the particle-method runs. 
Because we determined the broad dust density component with a Gaussian fit, where we could only use the outer wings of the dust density distribution, a lack of data in the midplane can introduce a certain level of uncertainty in the fit parameters, especially for the dust scale height. 
Correcting the fit parameters could lead to a smaller dust scale height across all setups.

The fact that both codes show a power-law trend in the width of the vorticity distribution with resolution (see Fig. \ref{fig: convergence rms vorticity}) indicates that this quantity has not yet saturated. 
We expect to see an even broader vorticity distribution with even larger values when the resolution is further increased. This could also mean that the vortex size increases.

\subsection{Comparison to other works}
\label{sec: comparison_to_other_work}
In Sect. \ref{sec: reconstruction_schemes} we computed dust scale heights for our particle simulations, which are on the order of 0.01 $H_{\rm g}$. 
These values are significantly larger than those reported in \cite{Flock2021} (they receive values close to 0.003 $H_{\rm g}$) for the same physical setup, but with a numerical issue that led to stronger settling. 
Other works, for instance \cite{LimSimon20252D}, report values of a similar order of magnitude to ours, indicating that we resolved the numerical issue in the current work. 

\cite{LimSimon20252D} have already done similar Lagrangian particle simulations using the ATHENA code in combination with a shearing box setup. 
When they increased the resolution, they also found a broadening of the midplane component, which is about to vanish in their highest-resolution run with $5120/H_{\rm g}$ (comparable to our simulation with $8192 \times 2048$). 
See also Fig. \ref{fig: athena_results} for a comparison of their vertical dust-to-gas ratios. 
The reason we find a midplane component for particles at the highest resolution in the current work, but they do not in the previous work, is that our domain size in the vertical direction is larger (0.4$H_{\rm g}$ vs. 0.2$H_{\rm g}$). 
The results in Sect. \ref{sec: influenceDomainSize} for our work and Fig. \ref{fig: athena_results_domain_size} for their work show that a larger vertical domain size leads to a thinner dust disk, which enhances the midplane component. 
Nevertheless, the results in \cite{LimSimon20252D} support that, if we increase the resolution further in particle simulations, we may see the same converged state for particles as for fluids, but this needs further checking.

In addition, \cite{LimSimon20252D} has shown that although increases in resolution broaden the midplane component, corresponding to a decrease in midplane dust density, it is beneficial in forming dense dust clumps that reach the critical Hill density (see their Sect. 3.4 for details). 
This could be because turbulence and dust-depleted voids become better resolved, as they can efficiently concentrate dust. 
The effect of this dust collection was already observed in \cite{YangZhu2021}, \cite{GerosaMeheut2023}, and recently studied in \cite{Umurhan2025}. 
The latter work connected the clumping to the turbulent concentration of particles. 
Based on our dust-fluid simulation results, this turbulence-driven dust concentration may be valid for both dust approaches.  

It is interesting that $H_{\rm d}$ of the midplane component increases with resolution, but not that of the broad component. 
It may be possible that each component traces a different turbulence-driven mechanism. 
For example, the midplane $H_{\rm d}$ could trace streaming turbulence, whereas the broad component could trace shear-driven instabilities, such as KHI, which should be most active near the top of the dust layer, if present. 
If this is true, it suggests that a critical resolution for convergence is higher for streaming turbulence. 
Because we find strong indications of convergence in the broad component, this raises the question of whether we can resolve the vertically shearing streaming instability \cite{Lin2021}. 
\cite{LimSimon20252D} already argued that this particular resolution may not be high enough. 
It would be necessary to further increase the resolution to determine whether we again observe a significant increase in the dust scale height for the broad density component. 

Our results of vanishing midplane density component raise another question: How do results from works like \cite{Li2021} or \cite{LimSimon20253D}, who constrained the parameter space in (total dust-to-gas mass ratio $Z$, Stokes number $\tau_s$), where disks are expected to show strong clumping, change with even higher resolutions? 
This issue was already addressed in Sect. 4.2 of \cite{LimSimon20253D}, which also includes a comparison with 3D simulations.

Our results of a vanishing midplane dust density component and a density structure covered with dust-depleted voids are consistent with the work of \cite{Yang2017}, who used the PENCIL code and Lagrangian particles. 
The reason might be their smaller domain size (0.2 $H_{\rm g}$ instead of our 0.4 $H_{\rm g}$, which can artificially increase the scale height, as shown in Sect. \ref{sec: influenceDomainSize}), or it might indicate that the midplane density component is merely a transient feature that disappears when the numerical schemes are sufficiently accurate or the (particle) resolution is sufficiently large. 
Assuming that this is a transient feature, we expect this component to vanish in PLUTO particle simulations as well, thereby demonstrating convergence. 
However, even at our highest resolution, we can still see this midplane component. 

The work of \cite{Baronett+2026Pfits} has carried out a comprehensive comparison of different hydrodynamical codes that treated dust as particles and as a fluid. 
For unstratified shearing-box simulations of streaming instability \citep[test BA from ][]{JohansenYoudin2007}, they found good agreement between the two approaches, where maximum dust densities for fluids are generally only 30$\%$ lower than for particles. 
We, however, find maximum dust densities to be at least 50$\%$ smaller for fluid simulations (comparing the maximum dust density for the highest resolution for the setups $\texttt{conf05}$ and $\texttt{conf07}$ in Table \ref{tab: overview}). 
Considering their resolution of 1024 cells per gas and dust scale height (due to an unstratified problem), we expect to find more similar values as we increase our resolution and expect a vanishing midplane component for particles. 
In Fig. \ref{fig: cumulativeDistribution}, we present the cumulative distribution function of the dust-to-gas ratio, indicating that once the midplane component vanishes, the probabilities are reduced significantly. 
Our plot can be compared to Fig. 12 in \cite{Baronett+2026Pfits}, for example.  

\subsection{Performance benchmark}
In Appendix \ref{sec: performance} we investigate the performance of fluid and particle simulations performed with FARGO3D (tested separately on central processing units (CPUs) and one graphics processing unit (GPU)) and PLUTO (tested on CPUs). 
The values presented in Table \ref{tab: performance} serve as a benchmark for the community and future works by providing order-of-magnitude estimates.  
The FARGO3D simulations had a factor of two higher vertical resolution. 
This should not affect the code performance too much because the full simulation time $\Delta t_{\rm Sim}$ and the resolution $N_{\rm r} \cdot N_{\theta}$ in Eq. \ref{eq: performance} are expected to increase by about a factor of 2. 
However, higher resolution can lead to longer data read times if more disk space is reserved for a simulation and the data are more spread out. 
This should reduce the performance to some degree, which we did not test here.

\subsection{Effect of numerical dissipation and limitations of the model}
We briefly comment on our choice of not including an explicit physical dissipation scale into our setups (in PLUTO, we did not use gas viscosity or dust diffusion at all, and in FARGO3D, the $\alpha$-viscosity and dust diffusion module was used with $\alpha$=0). 
When working with a fluid, it is recommended to account for a small, nonzero, and controllable scale of dissipation.
For gas, this might be the kinematic viscosity, for example, whereas for dust, which lacks viscosity, the analogous quantity is dust diffusivity. 
If this is not considered, the dissipation scale is entirely determined by the grid cell size and the related truncation errors of the numerical solution. 
Since we are working with a nonlinear system of equations, we cannot rule out the possibility that these truncation errors influence the solutions. 
This complicates our resolution comparison between FARGO3D and PLUTO, as grid-scale numerical artifacts could at some point dominate the density structure and quantities such as the dust-to-gas ratio. 
If a dissipation scale is considered, it needs to fulfill two aspects: first, it needs to be small enough not to influence the process of interest, in our case, the streaming instability. Second, the dissipation should have an impact at low and high resolutions. A combination of the two effects is typically difficult to achieve.

In addition, in our FARGO3D simulations, we enabled dust diffusion despite setting the $\alpha$-viscosity to zero.
Because the dust diffusivity was calculated using $\alpha$, we effectively got pure numerical diffusion from using the module. 
However, the numerical diffusion is expected to be very small as the dust diffusivity is identically zero.
To assess how much numerical diffusion affects the results, we have compared dust scale heights in Appendix \ref{sec: numericalDiffusivity}, where the dust diffusion module is not used in the lowest-resolution run. 
Within the level of variations, the results are the same.

In Figs. \ref{fig: dustDensity2D_ComparisonFluidParticle_Roe} and \ref{fig: dustScaleHeight_midDTGRatio_maxRhodust_FluidParticle_Roe}, we investigate the difference between PLUTO fluid and particle simulations for an increasing resolution. 
As we have stated at the beginning of this work, the purpose of our investigations is not to determine whether the differences in dust scale height and midplane dust-to-gas ratio arise from additional numerical diffusivity in the fluid approach or from the fundamental distinction that particle methods allow for crossing trajectories. 
Therefore, such tests need to be done in future work.

\section{Conclusion}
\label{sec: Conclusion}
We performed spherical 2D isothermal axisymmetric simulations of gas and dust in a global patch setup using the hydrodynamic codes PLUTO and FARGO3D. 
The goal was to investigate how dust structures that are formed by the nonlinear streaming instability generally depend on the dust treatment (fluid versus particles), resolution, and on the choice of the gas Riemann solver and spatial reconstruction scheme. 
The purpose was not to identify or quantify the pure effect of crossing trajectories in particle simulations, for example. 
Our results demonstrated that the choice of the numerical scheme and resolution significantly affects the evolution of the nonlinear streaming instability in stratified disk simulations. 

When we increased the resolution, the PLUTO fluid and FARGO3D simulations clearly converged toward a disk state described by a single-Gaussian dust density component with a dust scale height between 0.0129 $H_{\rm g}$ and 0.0141 $H_{\rm g}$ and a random field of small dust-depleted voids. 
At the highest resolution of $8192\times2048$, corresponding to a resolution of 5120/$H_{\rm g}$, the codes therefore agreed well (for fluid approach) for a resolution of 66 cells per dust scale height. 
At intermediate resolutions, an additional midplane component is visible. This transient feature disappeared for sufficiently high numbers of cells. 
As this additional component remained present in our highest-resolution simulation with PLUTO particles but shows an increasing dust scale height, we conclude that either an even higher grid resolution or more than $1.6 \cdot 10^{8}$ particles are necessary to achieve convergence toward the same density structure as in the fluid case.

When the resolution (due to either the grid or the particles) is continuously increased in a simulation using an HLL Riemann solver and a low-order reconstruction scheme, for example, we expect to see the midplane component at sufficiently high resolution. 
This feature will then disappear for an even higher number of grid cells to show convergence.  
In agreement with well-known concepts, a more accurate Riemann solver and/or a higher-order spatial reconstruction scheme can effectively reduce the necessary resolution so that a converged state is reached. 
However, higher-order numerical schemes alone do not guarantee convergence. 
In inviscid simulations in particular, there is no physical dissipation scale, so each grid refinement allows new, smaller scales to develop at the grid scale. 
These small-scale vortices can produce steeper gradients in density and vorticity, which might alter the structures of interest. 
All scales relevant to a given structure must therefore be resolved first. 
Only then can higher-order schemes effectively reduce numerical errors and yield converged results for these scales. 

When we compared PLUTO fluid and FARGO3D simulations, we found that the FARGO3D dust density structure is comparable to a more accurate setup of PLUTO, using a ROE Riemann solver and a fourth-order reconstruction scheme, for example. 
However, FARGO3D consistently reported larger dust scale heights. 
Compared to PLUTO, we showed that the gas velocity distribution at small resolution is much closer to the distribution at the highest resolution 

In general, the differences we found between the two approaches are consistent with other code-comparison work, such as \cite{Baronett+2026Pfits}, and they are smaller than the uncertainties in the true Stokes number of dust grains, for instance. 
For the average dust-to-gas ratio in the midplane at the highest resolution, the PLUTO fluid and particle results are only 1.3 and 1.9 times the FARGO3D result, respectively. 
For the maximum dust density, the corresponding factors are 1.2 (PLUTO fluid) and 2.8 (PLUTO particles) relative to FARGO3D. 
Because the fluid approach can be faster by 3.5 times at least than the particle approach, we can use it as a viable method for studying the streaming instability in global disk setups.

\begin{acknowledgements}
This research has benefited from the PyPLUTO package for data analysis and visualization \citep[see][]{PyPLUTO2025}. In addition, we thank David Melon Fuksman, Prakruti Sudarshan, Alexandros Ziampras, Vittoria Berta, Kengo Tomida, Stanley Baronett, Pinghui Huang, Xue-Ning Bai, Yu Wang, Min-Kai Lin, Hubert Klahr, Orkan Umurhan, and Pablo Benítez-Llambay for helpful discussions. We thank Tilman Dannert from the Max Planck Computing and Data Facility (MPCDF) team for guidance on setting up FARGO3D simulations on GPUs of the MPCDF cluster Vera. In addition, we want to thank the referee and language editors for helpful feedback and suggestions for improving this work.
\end{acknowledgements}

\bibliographystyle{bibtex/aa.bst}
\bibliography{bibtex/referencePaper}

\begin{appendix}
\nolinenumbers

\section{Impact of the spatial reconstruction order and gas Riemann solver on gas vorticity}
\label{sec: SpatialRiemannVorticity}
\begin{figure}[ht]
    \centering
    \includegraphics[]{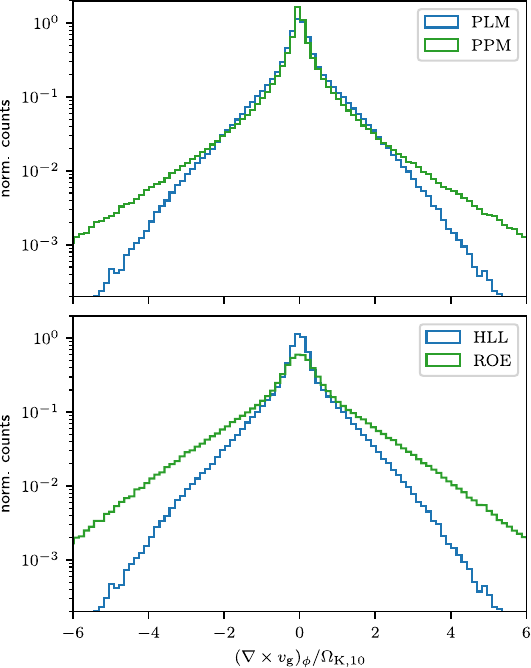}
    \caption{Distribution of the azimuthal component of the gas vorticity for PLM and PPM (\textit{top}) as well as HLL and ROE (\textit{bottom}) after 100 orbits of evolution. $\Omega_{\rm K, 10}$ is the Keplerian angular velocity at 10 AU.}
    \label{fig: VorticityRiemannSolver_SpatialRecon}
\end{figure}

To better understand the reason why a higher-order reconstruction scheme as well as a more accurate Riemann solver lead to an additional midplane component in the dust density, as seen in Sects. \ref{sec: reconstruction_schemes} and \ref{sec: riemann_solver}, we compare the gas vorticity component in the azimuthal direction for the two reconstruction schemes PLM and PPM (top panel) and solvers HLL and ROE (bottom panel) in Fig. \ref{fig: VorticityRiemannSolver_SpatialRecon}. 
The results indicate that both PPM and ROE lead to more frequent higher vorticity values. 
In addition, since the distribution for PLM and HLL are identical, we also see that ROE leads to more frequent and larger values in the range $(\nabla \times v_{\rm g})_{\phi}$ = [-2, 2] $\Omega_{\rm K, 10}$ compared to PPM.

\section{Disk mass evolution for particle and fluid simulations}
\label{sec: dustDiskMass}
\begin{figure}[ht]
    \centering
    \includegraphics[]{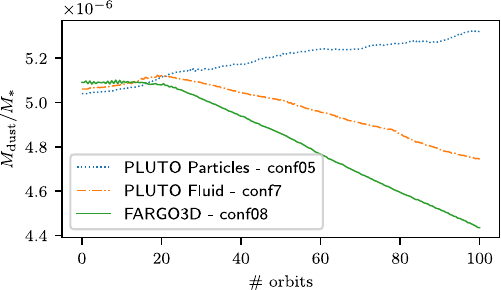}
    \caption{Temporal evolution of the dust disk mass for more accurate PLUTO (fluid, particles) simulations as well as a FARGO3D simulation. 
    All simulations have a resolution of $4096\times1024$. 
    The radial domain used to compute the dust mass is limited from 9.4 to 10.6 AU, excluding the damping zone where dust is replenished.}
    \label{fig: dustDiskMasses}
\end{figure}

Figure \ref{fig: dustDensity2D_ComparisonFluidParticle_Roe} shows that not enough dust was delivered at the outer boundary for the particle simulation, as in the midplane, the dust density component disappears. 
However, for particle simulations, we did not see this drop in dust density. 
Therefore, we test here how the total dust mass varies over time for the most accurate PLUTO particle and fluid simulations, as well as for a FARGO3D simulation. 
For all three setups, we chose only the setup with a resolution of $4096\times1024$. The results are shown in Fig. \ref{fig: dustDiskMasses}.  

The results indicate that, in particle simulations, the mass increases over time and that excessive dust is added at the outer boundary. 
For fluid simulations, however, not enough mass is indeed added at the outer boundary. 

\section{Influence of numerical diffusivity for FARGO3D runs due to the diffusion module}
\label{sec: numericalDiffusivity}
\begin{figure}[ht]
    \centering
    \includegraphics[]{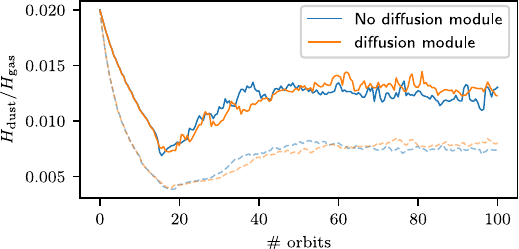}
    \caption{Comparison of the dust scale height for FARGO3D simulations with a resolution of 512$\times$128. 
    The orange lines show the dust scale heights for a simulation where the dust diffusion module is used with a value of zero for the turbulent $\alpha$-parameter, while the blue lines show the dust scale height for a simulation where no diffusion module is used. 
    Dark, solid lines indicate the dust scale height of the broad dust density component,
    while light, dashed lines correspond to the midplane density component.}
    \label{fig: ComparisonDiffusionActive_NoDiffusionActive}
\end{figure}

As mentioned in Sect. \ref{sec: numericalSetups}, we used the diffusion module of FARGO3D but chose $\alpha = 0.0$ for the strength of turbulence.  
We compared a simulation with a resolution of $512\times128$ and dust diffusion activated ($\alpha=0$) to a simulation without dust diffusion. 
This serves as a test of whether, and by how much, numerical diffusion, combined with a resolution of about 4 cells per dust scale height, affects the dust scale height.
We compare the dust scale height only in the radial domain of 10 $\pm$ 0.15 AU. 
The result in Fig. \ref{fig: ComparisonDiffusionActive_NoDiffusionActive} shows that within the fluctuations, both profiles are identical. 

\section{Origin of different dust scale heights between FARGO3D and PLUTO at low resolution}
\begin{figure}[ht]
    \centering
    \includegraphics[]{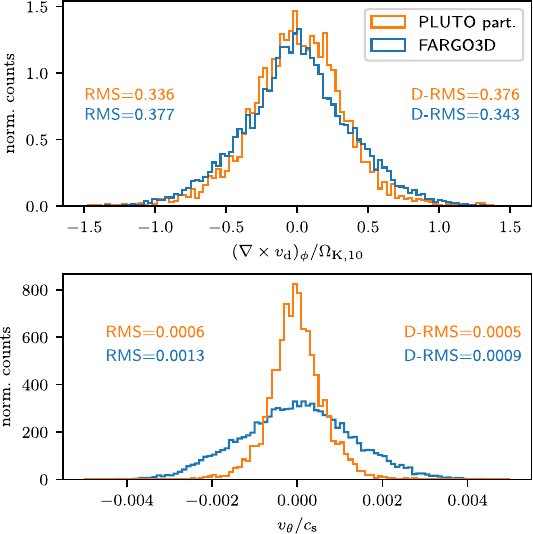}
    \caption{\textit{Top}: Distribution of vorticity for PLUTO particles (orange) and FARGO3D (blue). 
    \textit{Bottom}: The meridional dust velocity distribution for the same two setups. 
    The values on the left-hand side in each subplot show the standard RMS values, while the values on the right-hand side show the density-weighted RMS values.
    The data were taken after 50 orbits of disk evolution for a resolution of 512$\times$128.
    Both distributions considered only data between 9.4 and 10.4 AU in the radial direction and between -0.05 and 0.05 AU in the vertical direction.}
    \label{fig: vorticity velocity lowest res}
\end{figure}

In this section, we briefly look at the distribution of dust vorticity and meridional dust velocity at the lowest resolution (512$\times$128 for FARGO3D and PLUTO particles at 50 orbits of evolution. 
We computed the standard RMS, as well as the density-weighted root-mean square (D-RMS; see also Fig. \ref{fig: convergence rms vorticity}) for both quantities and show it in Fig. \ref{fig: vorticity velocity lowest res}. 
The RMS value gives us the width of the distribution, while the D-RMS focuses on regions with high dust density, for example, around the midplane.  
For the particle simulation, we did not consider vorticity or velocity that is exactly zero, as these are associated with regions not described by particles.

The results for the vorticity indicate that at the lowest resolution, the density-weighted RMS value for PLUTO particles is roughly 10$\%$ higher compared to FARGO3D's value.
In the regions where most of the dust is located, both codes show similar vorticity distributions. 
For the meridional dust velocity, we can see that, regardless of how the distribution width is measured (standard root-mean-square or density-weighted), FARGO3D yields a distribution that is twice as broad as that of PLUTO particles.
Therefore, FARGO3D shows higher bulk velocities in the meridional direction.

\section{Distribution of particles per cell}
\label{sec: particles_per_cell}
\begin{figure}[ht]
    \centering
    \includegraphics[]{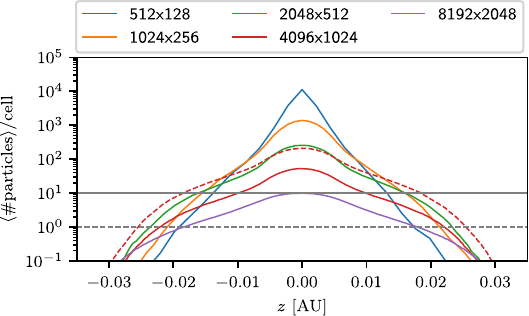}
    \caption{Vertical distribution of the average number of particles per cell for different resolutions. 
    The horizontal dashed~gray line marks the boundary to values smaller than 1 particle per cell, while the solid line marks 10 particles per cell. 
    The red dashed line shows the vertical distribution when we increase the number of particles to $4.16\cdot 10^{7}$.}
    \label{fig: particlesPerCell}
\end{figure}

As we fixed the total number of particles across different resolutions for setup \texttt{conf05}, the average number of particles per cell in the vertical direction changes. 
At the lowest resolution, we expect many more particles near the midplane than at the highest resolution, which may yield a more accurate description. 
We tested whether the average number of particles per cell in the vertical direction affects the vertical thickness of the dust disk, for instance, when the location $z$ above the midplane where the dust density starts to drop more strongly with height coincides with the height where the average number of particles reaches a value of 1. 
We proceed as follows: For each resolution, we considered only simulation data from 30 orbits onward and only particles located between 9.4 and 10.6 AU. 
These particles were then binned into a number of $N_{\theta}$ bins according to their polar coordinate. 
This way, the binning is independent of the radial location. This particle-location distribution was then averaged over time. 
Finally, we transformed the distribution from polar to height above the midplane, using a radius of 10 AU, which is shown in Fig. \ref{fig: particlesPerCell}.

A few results can be pointed out here: Firstly, for a resolution of $1024\times256$ and higher, when going from the midplane to higher locations, the distribution shows a stronger decrease at around $z\approx0.02$AU, which can also be seen in the third row of Fig. \ref{fig: dustDensity2D_ComparisonFluidParticle_Roe}. 
This stronger decrease occurs at roughly 10 particles per cell at a resolution of $1024\times256$, and at a smaller number at higher resolution. 
Secondly, for a resolution of $2048\times512$, we can see that for values below one particle per cell, the distribution shows an even stronger decrease. 
This behavior was already shown and observed in Fig. A1 of \cite{Flock2021}. 
For the two highest-resolution runs, the location where we find around 1 particle per cell lies at 0.02 AU. 
Finally, if we change the number of particles from $1.04\cdot 10^{7}$ to $4.16\cdot 10^{7}$ for a resolution of $4096\times1024$ (red dashed line), the height at which the average number of particles drops below a value of 1 increases from 0.0225 AU to 0.026 AU.

\section{Cumulative distribution function for dust-to-gas ratios}
As in \cite{Benítez-Llambay2019} (see their Fig. 10) or \cite{Baronett+2026Pfits} (see their Fig. 12), we can compute the cumulative distribution function for the dust-to-gas ratio. 
We only considered values in the radial domain from 9.4 to 10.4 AU. 
To obtain this distribution function, we considered simulation outputs from 30 to 100 orbits (in steps of 2.5 orbits) for a resolution of $4096\times1024$ and from 30 to 50 orbits (in steps of one orbit) for a resolution of $8192\times2048$. 
This ensures that we consider both a spatial and temporal variation. 
Each distribution was normalized so that the probability of finding a value of 0.01 was exactly 1. 
\begin{figure}[ht]
    \centering
    \includegraphics[]{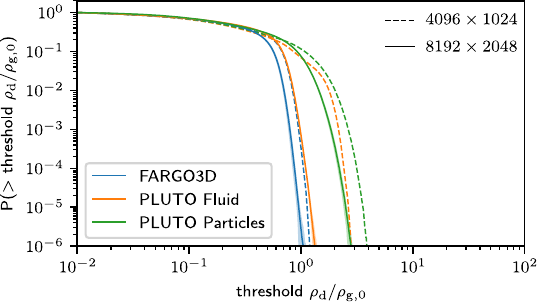}
    \caption{Cumulative distribution function for the three setups: FARGO3D (\texttt{conf08}), PLUTO fluids (\texttt{conf07}), and PLUTO particles (\texttt{conf05}). 
    The solid line shows the mean, while the shaded area around the solid line indicates one standard deviation.}
    \label{fig: cumulativeDistribution}
\end{figure}

Comparing our results with Fig. 12 in \cite{Baronett+2026Pfits}, for example, we find overall smaller values (roughly one order of magnitude lower). 
This can be due to the different setups (global stratified simulations in the current work and non-stratified shearing box simulations in their work) as well as due to a different resolution of the dust (In the highest resolution case, we describe our dust with roughly 70 cells per dust scale height, while they have 1024 cells due to a non-stratified disk).

When the resolution changed from $4096\times1024$ to $8192\times2048$, the probabilities for each setup decreased. 
For PLUTO fluid and particle simulations, this is most likely linked to the broadening of the midplane component. 
Because this component is completely absent in PLUTO fluids, we observe a stronger reduction in probabilities than for particles where the component remains visible. 
For FARGO3D, the reduction in probabilities is linked to a broader dust scale height, which spreads dust over a larger vertical domain, thereby reducing the dust-to-gas ratio.

\section{Vertical dust-to-gas ratio profile obtained with the ATHENA code}
\label{sec: athena}
\begin{figure}[ht]
    \centering
    \includegraphics[width=\columnwidth]{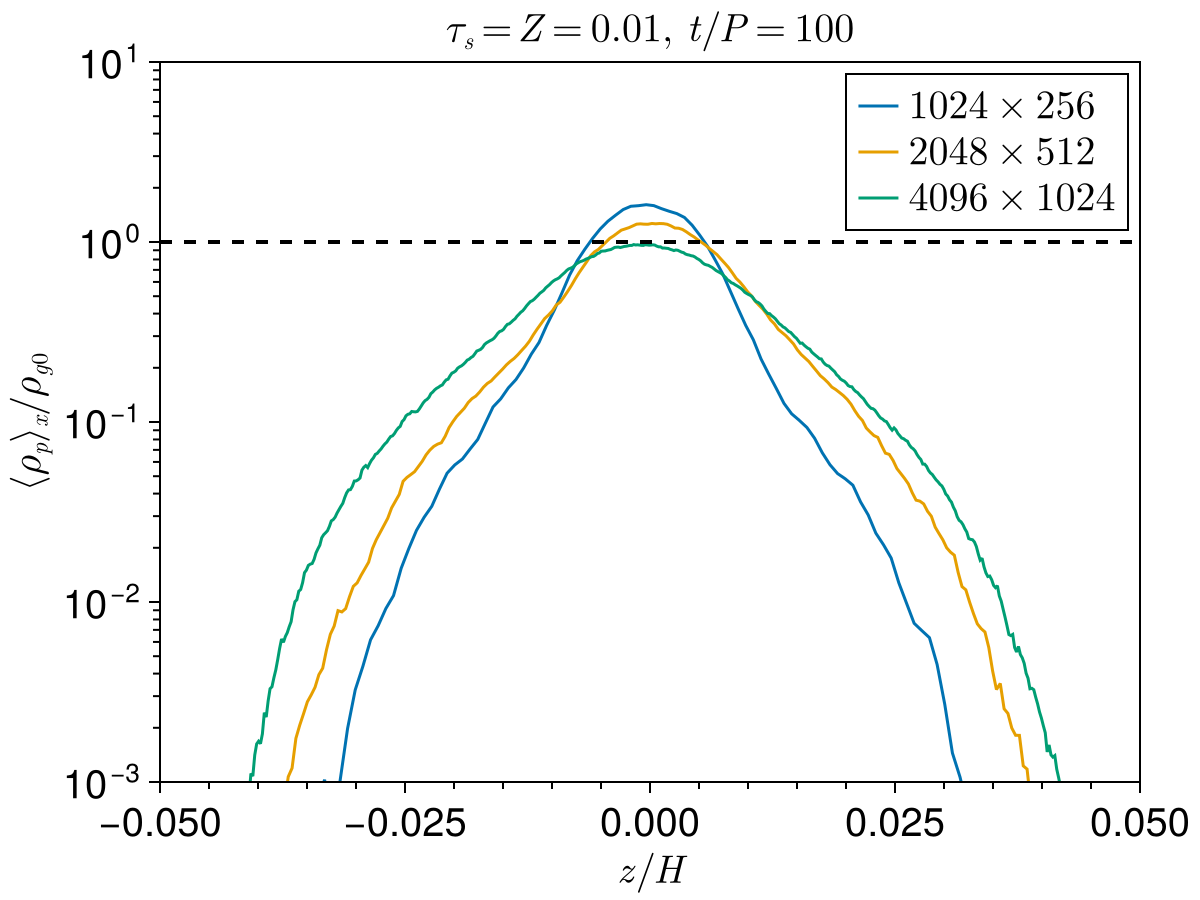}
    \caption{Vertical dust-to-gas ratio profile for the three simulations in Fig. 6 of \cite{LimSimon20252D}, which were obtained with ATHENA and Lagrangian particles. 
    The three profiles with $1024\times256$, $2048\times512$, and $4096\times1024$, correspond to a number of 1280/$H_{\rm g}$, 2560/$H_{\rm g}$, and 5120/$H_{\rm g}$, respectively. 
    The dashed~black line indicates a dust-to-gas ratio of unity.}
    \label{fig: athena_results}
\end{figure}

Figure \ref{fig: athena_results} presents a vertical dust-to-gas ratio profile for the three simulations shown in Fig. 6 of \cite{LimSimon20252D}. 
These simulations were performed using Lagrangian particles and a full vertical domain of 0.2$H_{\rm g}$; they show the disk state after 100 orbits, and the profiles were obtained by averaging over the entire radial domain. 
This result shows that the midplane component for their highest resolution of 5120/$H_{\rm g}$ is about to vanish and is therefore even broader than the component in the top row of Fig. \ref{fig: dustDensity2D_ComparisonFluidParticle_Roe} for a resolution of $8192\times2048$, which corresponds to the same number of cells per gas scale height.

\begin{figure}[ht]
    \centering
    \includegraphics[width=\columnwidth]{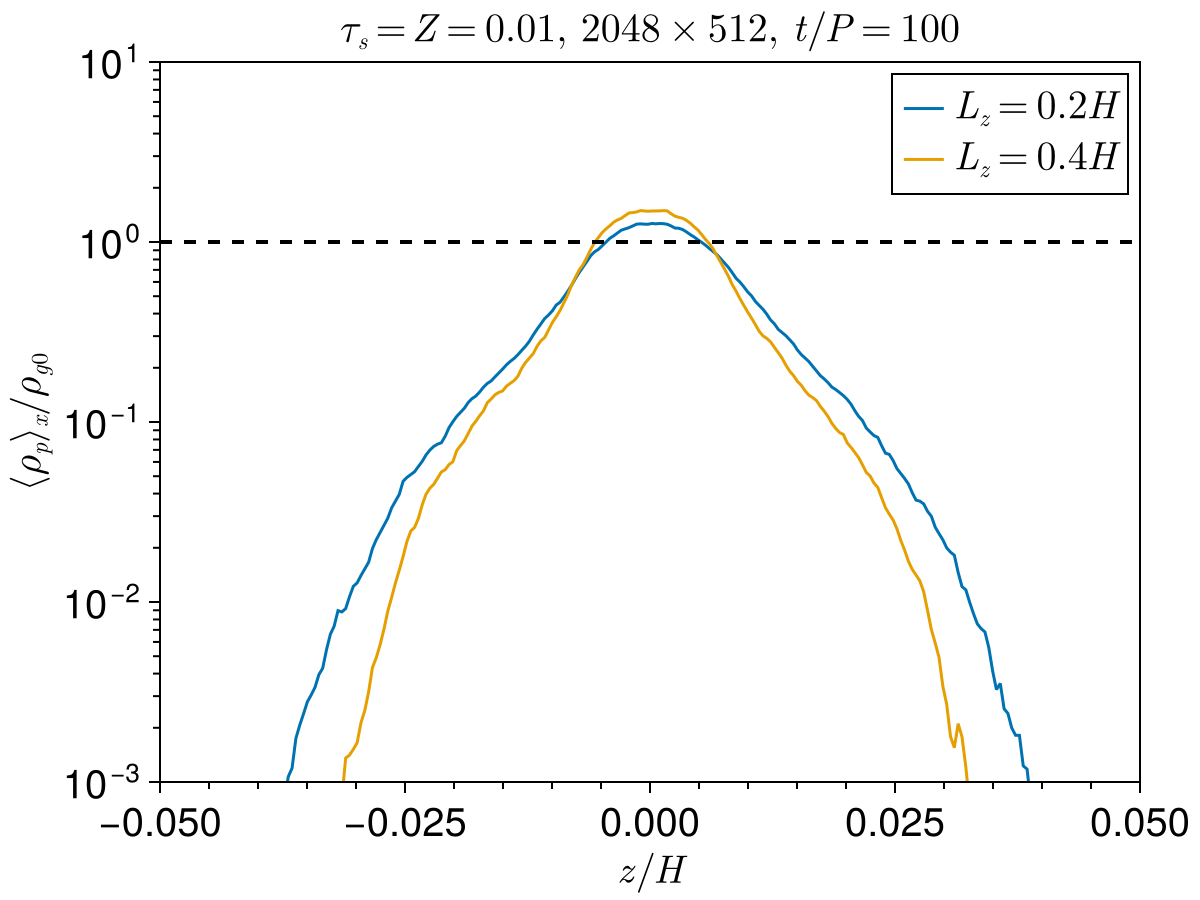}
    \caption{Vertical dust-to-gas ratio profile for a resolution of $2048\times 512$ for two vertical domain sizes. 
    The dashed black line indicates a dust-to-gas ratio of unity.}
    \label{fig: athena_results_domain_size}
\end{figure}

They also performed a test to assess how much the vertical domain size influences the resulting dust scale height (see their Sects. 2.2.2 and 4.2.2). 
We plot in Fig. \ref{fig: athena_results_domain_size} the vertical dust-to-gas ratios for a resolution of $2048 \times 512$ when they increase the vertical domain size to 0.4$H_{\rm g}$.

\newpage
\section{Disk structure due to dust-depleted voids}
\label{sec: depleted voids}
\begin{figure}[ht]
    \centering
    \includegraphics[]{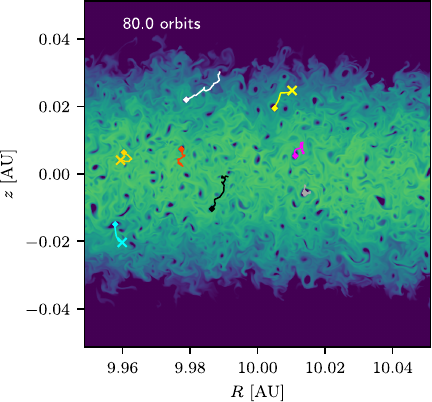}
    \caption{Trajectories of 8 vortices over a time span of 1 orbit. 
    The dust structure in the background is from the FARGO3D simulation at a resolution of $8192\times2048$ and shows the dust density after 80 orbits. 
    The diamond-shaped markers indicate the initial position at 80 orbits, while the crosses indicate that a vortex has dissolved and its location could not be detected.}
    \label{fig: void trajectories}
\end{figure}

In this section we briefly show that dust-depleted voids, which are clearly visible in the highest-resolution simulations of FARGO3D and PLUTO fluid, are short-lived features. 
We let the FARGO3D simulation, with resolution $8192\times2048$, continue and produced outputs after 0.05 orbits over a time span of 2.75 orbits. 
First, we examined the trajectories of 8 different vortices and tracked their positions over 1 orbit. 
We determined the location of these voids by hand and always tried to find a suitable center. 
Some voids became elongated, making it difficult to set a location. 
If we couldn't determine a clear center, we said the void dissolved, and we did not try to track future positions. 
The result is shown in Fig. \ref{fig: void trajectories}. 
Vortices near the midplane move slowly and circulate around their initial positions. 
Instead, vortices at higher altitudes can migrate in a given direction (either closer to the midplane or toward the upper and lower disk edge). 

\begin{figure}[ht]
    \centering
    \includegraphics[]{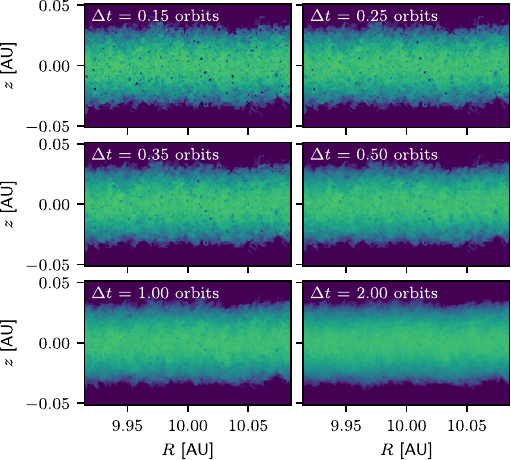}
    \caption{Temporal average of the dust density structure over various time windows of size $\Delta t$.}
    \label{fig: temporal average vortices}
\end{figure}

In Fig. \ref{fig: temporal average vortices}, we show how the dust density structure looks when performing temporal averages over various time windows, always starting at 80 orbits. 
For instance, the average for $\Delta t$=0.15 is computed using 3 outputs. 
For a time window of half an orbit, most vortices have disappeared already, and only a few locations near the midplane are visible where the average dust density is still reduced due to vortices. 
These locations have disappeared after a 2-orbit window. 
Therefore, even slowly moving vortices do not live longer than 2 orbits.

\section{Code performance}
\label{sec: performance}
In Table \ref{tab: performance} we provide an overview of performance across different numerical setups. 
This includes the most accurate and least accurate setups for the fluid and particle approaches using PLUTO, as well as FARGO3D simulations. 
A measurement for the performance is the number of cell updates per second, which we describe as the performance metric $\rm Perf$, and which can be computed in the following way
\begin{equation}
\label{eq: performance}
    \mathrm{Perf} = \frac{N_{\rm cells, tot}}{\Delta t_{\rm Sim}} = \frac{N_{\rm dt} \cdot N_{\rm r} \cdot N_{\theta}}{\Delta t_{\rm Sim}}.
\end{equation}
Here, $N_{\rm cells, tot}$ denotes the total number of cell updates performed during the simulation time $\Delta t_{\rm Sim}$. It is given by the number of time steps, $N_{\rm dt}$, multiplied by the number of cells per time step, $N_{\rm r} \cdot N_{\theta}$.
Here, $N_{\rm r}$ and $N_{\theta}$ represent the number of cells in the radial and meridional directions, respectively. 

To receive performance measurements, which allow a reliable comparison between codes and setups, each simulation with a resolution smaller than $8192\times2048$ was run on 16 CPUs. 
For FARGO3D, we added one simulation with resolution $8192\times2048$, which was run on a single GPU. 
Each simulation was run for a different number of orbits (at 10 AU); however, we determined the number of time steps per orbit. 
The smallest number of time steps was 25500 in the lowest-resolution FARGO3D run, which should be sufficient to provide a good performance estimate.

Three different computing clusters were employed, as shown in the hardware column (HW). 
VP and VA denote the MPCDF computing clusters VIPER\footnote{\url{https://docs.mpcdf.mpg.de/doc/computing/viper-user-guide.html}} and VERA\footnote{\url{https://docs.mpcdf.mpg.de/doc/computing/clusters/systems/Astronomy.html}}, respectively, while AN represents an in-house cluster dedicated to local computations. 
The CPU models and compilers used across the clusters vary; therefore, we list them here for full transparency. 
On the in-house computing cluster (AN), we used Intel(R) Xeon(R) Gold 6132 CPUs and compiled every code with Intel oneAPI DPC++/C++ (mpiicx, version 2021.11), including -O3 -ipo -qopt-mem-layout-trans=4 -fimf-use-svml=true and linked with the corresponding Intel MPI and math libraries; the parallel build used the PARALLEL preprocessor configuration. 
On Vera (VA), Intel(R) Xeon(R) Platinum 8360Y CPUs were installed. 
We used Intel oneAPI DPC++/C++ (mpiicx, version 2021.11) together with the -O3 -march=znver4 optimization and linked against the standard MPI and math libraries. 
For computations on a single GPU, we used an NVIDIA A100-SXM4-40GB. 
The code was compiled with NVIDIA nvcc (version 12.1.1) using the -03 -arch=sm\_80 optimization flag, and linked with the CUDA runtime (-lcudart) and the standard C++ libraries. 
Finally, we used AMD EPYC 9554 CPUs on Viper (VP), where the code is compiled with Intel oneAPI DPC++/C++ (mpiicx version 2021.11) with optimization flags -O3 -march=znver4 for both compilation (-c) and linking, against the MPI library (-lmpi). 

From the performance results in Table \ref{tab: performance}, we observe the following: 
Compared to the highest-resolution PLUTO fluid simulations, the PLUTO particle simulations at the same resolution show a performance reduction by a factor of 3.5–4.
For dust-fluid simulations, the number of cells that are advanced in time is twice the given resolution; at a resolution of $4096\times512$, for instance, we have approximately $4\cdot10^{6}$ cells. 
For a dust particle simulation at the same resolution, we have approximately $2\cdot10^{6}$ gas cells and $1\cdot10^{7}$ particles to evolve. 
Therefore, the number of elements to evolve for particle simulation is roughly a factor of 3 higher than for pure fluid simulations. 
Since the number of particles was held fixed while the grid resolution was varied, the performance gap between particle-based and pure-fluid PLUTO simulations widens at lower resolutions. 
This is because the particle update per time step, $\Delta t$, dominates the computational cost and occurs at a fixed rate (total particle updates per second), independent of resolution. 
As each update advances the entire gas grid once, the performance in $\mathrm{cells\,s^{-1}}$ scales with the total number of grid cells times the particle update rate.
This value drops at low resolutions (e.g., $512\times64$) since few cells are updated per second. 
In contrast, in pure fluid simulations, cells advance by $\Delta t$, so the single-cell update rate directly determines the maximum number of cells that can be updated in one second. 

Furthermore, by comparing the performance of the most accurate (\texttt{conf07}) and least accurate (\texttt{conf06}) PLUTO fluid setups on the AN cluster, we observe that the more accurate setup is approximately 1.6 times slower than the least accurate one. 
This performance loss can be mitigated by using a newer-generation CPU cluster (see results for \texttt{conf07} on the VP cluster at the highest resolution). 
For the most and least accurate setups with particles, \texttt{conf05} and \texttt{conf01}, respectively, the performance loss by switching to a more accurate setup is low (see a performance of 2.19 $\cdot 10^{6}$ $\mathrm{cells\,s^{-1}}$ for setup \texttt{conf05} and 2.26 $\cdot 10^{6}$ $\mathrm{cells\,s^{-1}}$ for setup \texttt{conf01}). 
This is primarily because the numerical setup difference mainly applies to the gas, which is numerically more efficient than handling many particles. 
However, with the newer generation of CPUs on VP, we get an additional 2x speedup.

Finally, we compare the performance of pure hydrodynamic simulations using FARGO3D and PLUTO (\texttt{conf07}). For resolutions below $8192\times2048$, both codes show a comparable performance on CPUs, with FARGO3D only 7–14$\%$ faster when using the AN cluster for PLUTO and VA for FARGO3D. Migrating PLUTO from the AN cluster to the newer VP cluster yields a 54$\%$ performance increase. For comparison, a FARGO3D run on a single GPU at $8192\times2048$ achieves an additional 32$\%$ speedup relative to CPU-based runs.

The results therefore show that both hydrodynamics codes, using the fluid approach for dust, have similar performance.
By switching to a newer version of CPUs, we can even achieve a performance that is higher than that on a GPU\footnote{We are aware that the current number of cells does not fully saturate the GPU memory as only 7GB of 40GB are used, and we would expect a higher performance when switching to 3D models.}

The benefit of computing the $\# \rm dts/$orbit is that the impact of different CFL values used during the computation is directly visible. 
If we focus on the PLUTO simulations alone, we can observe that both more and less accurate simulations of a dust fluid (\texttt{conf06} and \texttt{conf07}), as well as more accurate simulations of dust particles (\texttt{conf05}), yield similar values at a given resolution, since all were run with a CFL value of 0.2. 
FARGO3D simulations (which use a CFL value of 0.44) show that only half of the $\# \rm dts/$orbit at a given resolution are necessary compared to a CFL of 0.2.  
\begin{table}
\caption{Performance comparison for the different dust modeling approaches and both hydrodynamics codes.}
\label{tab: performance}
\centering
\begin{tabular}{l c c c}
\hline\hline
Resolution   &    HW        & $\#$dts/orbit  & $\rm Perf$   \\ 
             &              &                & [$\#$cells/s]   \\ 
\hline \\ [-1.0ex]

\multicolumn{4}{c}{PLUTO Particles - \texttt{conf01}} \\
\hline 
    4096x512 &   AN         & 78302   &  \pgfmathprintnumber{2260052} \\
\\ [-1.0ex]

\multicolumn{4}{c}{PLUTO Particles - \texttt{conf05}} \\
\hline
    512$\times$64   &   AN         & 9794    &  \pgfmathprintnumber{33578}   \\
    1024$\times$128 &   AN         & 19602   &  \pgfmathprintnumber{100483}  \\
    2048$\times$256 &   AN         & 39260   &  \pgfmathprintnumber{492558}  \\
    4096$\times$512 &   AN         & 78491   &  \pgfmathprintnumber{2194656} \\
    4096$\times$512 &   VP         & 78378   &  \pgfmathprintnumber{4011021} \\
\\ [-1.0ex]

\multicolumn{4}{c}{PLUTO Fluid - \texttt{conf06}} \\
\hline
    4096$\times$512 &   AN         & 78278   &  \pgfmathprintnumber{14367358} \\
\\ [-1.0ex]

\multicolumn{4}{c}{PLUTO Fluid - \texttt{conf07}} \\
\hline
    512$\times$64   &   AN         & 9791    &  \pgfmathprintnumber{8120476}  \\
    1024$\times$128 &   AN         & 19603   &  \pgfmathprintnumber{8850150}  \\
    2048$\times$256 &   AN         & 39272   &  \pgfmathprintnumber{8934030}  \\
    4096$\times$512 &   AN         & 78679   &  \pgfmathprintnumber{9099663}  \\
    4096$\times$512 &   VP         & 78376   &  \pgfmathprintnumber{14088547} \\
\\ [-1.0ex]

\multicolumn{4}{c}{FARGO3D - \texttt{conf08}} \\
\hline
    512$\times$128   &   VA         & 5100   &  \pgfmathprintnumber{8704000}  \\
    1024$\times$256  &   VA         & 10190  &  \pgfmathprintnumber{9720696}  \\
    2048$\times$512  &   VA         & 20390  &  \pgfmathprintnumber{10065184} \\
    4096$\times$1024 &   VA         & 40770  &  \pgfmathprintnumber{10312370} \\
    8192$\times$2048 &   VA (1 GPU) & 81530  &  \pgfmathprintnumber{13585809} \\
\hline\hline
\end{tabular}
\tablefoot{The second column lists the different hardware (HW) components or high-performance clusters that were used for simulations. AN stands for a computing cluster located in the institute, VP for a Viper, and VA for Vera. For more information about the clusters and the models used for CPU and GPU, see Appendix \ref{sec: performance}.}
\end{table}

\end{appendix}

\end{document}